\documentclass{emulateapj}
\usepackage{longtable}

\newcommand{\cote}{C\^{o}t\'{e}\ }
\newcommand{\jordan}{Jord\'{a}n\ }
\newcommand{\etal}{et~al.\ }
\newcommand{\gz}{($g$--$z$)}
\newcommand{\ggzz}{($g_{475}$--$z_{850}$)}

\slugcomment{Accepted for publication in the Astrophysical Journal}

\shorttitle{Color Distributions of Globular Cluster Systems}
\shortauthors{Peng et al.}

\begin{document}


\title{The ACS Virgo Cluster Survey IX: The Color Distributions of
  Globular Cluster Systems in Early-Type Galaxies\altaffilmark{1}}


\author{Eric W. Peng \altaffilmark{2,3,4}}

\author{Andr\'{e}s Jord\'{a}n \altaffilmark{5,6}}

\author{Patrick C\^{o}t\'{e} \altaffilmark{2}}

\author{John P. Blakeslee \altaffilmark{7}}

\author{Laura Ferrarese \altaffilmark{2}}

\author{Simona Mei \altaffilmark{7}}

\author{Michael J. West \altaffilmark{4,8}}

\author{David Merritt \altaffilmark{9}}

\author{Milos Milosavljevi\'{c} \altaffilmark{10,11}}


\author{John L. Tonry \altaffilmark{12}}

\altaffiltext{1}{Based on observations with the NASA/ESA {\it Hubble
    Space Telescope} obtained at the Space Telescope Science Institute,
    which is operated by the Association of Universities for Research in
    Astronomy, Inc., under NASA contract NAS 5-26555.}
\altaffiltext{2}{Herzberg Institute of Astrophysics, 
  National Research Council of Canada, 
  5071 West Saanich Road, Victoria, BC  V9E 2E7, Canada; 
  Eric.Peng@nrc-cnrc.gc.ca, Patrick.Cote@nrc-cnrc.gc.ca, 
  Laura.Ferrarese@nrc-cnrc.gc.ca}
\altaffiltext{3}{Department of Physics and Astronomy, Rutgers
  University, New Brunswick, NJ 08854, USA}
\altaffiltext{4}{Visiting Astronomer, Cerro Tololo Inter-American
    Observatory, which is operated by the Association of Universities
    for Research in Astronomy, Inc. (AURA), under cooperative
    agreement with the National Science Foundation.}
\altaffiltext{5}{European Southern Observatory, 
  Karl-Schwarzschild-Str. 2, 85748
  Garching bei M\"{u}nchen, Germany; ajordan@eso.org}
\altaffiltext{6}{Astrophysics, Denys Wilkinson Building, University of
  Oxford, 1 Keble Road, OX1 3RH, UK}
\altaffiltext{7}{Department of Physics and Astronomy, 
  Johns Hopkins University, Baltimore, MD 21218, USA;
  jpb@pha.jhu.edu, smei@pha.jhu.edu}
\altaffiltext{8}{Department of Physics and Astronomy, University of Hawaii, 
  Hilo, HI 96720, USA; westm@hawaii.edu}
\altaffiltext{9}{Department of Physics, Rochester Institute of Technology,
Rochester, NY 14623-5604, USA; merritt@astro.rit.edu}
\altaffiltext{10}{Theoretical Astrophysics, California Institute of 
  Technology, 
  Mail Stop 130-33, Pasadena, CA 91125, USA; milos@tapir.caltech.edu}
\altaffiltext{11}{Sherman M. Fairchild Fellow}
\altaffiltext{12}{Institute for Astronomy, University of Hawai'i, 2680 Woodlawn
  Drive, Honolulu, HI 96822, USA; jt@ifa.hawaii.edu}

\begin{abstract}

We present the color distributions of globular cluster (GC) 
systems for 100 Virgo
cluster early-type galaxies observed in the ACS Virgo Cluster
Survey, the deepest and most homogeneous survey of this kind to date.
While the color distributions of individual GC systems can
show significant variations from one another, their general properties
are consistent with continuous trends across galaxy luminosity, color,
and stellar mass.  On average, galaxies at all luminosities in our study
($-22 < M_B < -15$) appear to have bimodal or asymmetric GC color
distributions. Almost all
galaxies possess a component of metal-poor GCs, with the average
fraction of
metal-rich GCs ranging from 15 to 60\%. 
The colors of both subpopulations correlate with
host galaxy luminosity and color, with the red GCs having a steeper
slope.  The steeper correlation seen in the mean color of the entire GC
system is driven by the increasing fraction of metal-rich GCs for more
luminous galaxies.

To convert color to metallicity, we also introduce a preliminary
\gz-[Fe/H] relation calibrated to Galactic, M49 and M87 GCs.  This
relation is nonlinear with a steeper 
slope for ${\rm [Fe/H]} \lesssim -0.8$.  As a result, the metallicities
of the metal-poor and metal-rich GCs vary similarly with respect to
galaxy luminosity and stellar mass, with relations of 
${\rm [Fe/H]}_{MP}\propto L^{0.16\pm0.04}\propto M_{\star}^{0.17\pm0.04}$ and
${\rm [Fe/H]}_{MR}\propto L^{0.26\pm0.03}\propto M_{\star}^{0.22\pm0.03}$, 
respectively.   Although these relations are shallower than the
mass-metallicity relation predicted by wind models 
and observed for dwarf galaxies,
they are very similar to the mass-metallicity relation for star forming
galaxies in the same mass range.  
The offset between the two GC populations varies slowly ($\propto
M_\star^{0.05}$) and is approximately 1~dex across three orders of
magnitude in mass, suggesting a nearly universal amount of enrichment between
the formation of the two populations of GCs.  We also find that although
the metal-rich GCs show a larger dispersion in color, it is the
{\it metal-poor GCs} that have an equal or larger dispersion in metallicity.
The similarity in the $M_\star$--[Fe/H] relations for the two
populations, implies that
the conditions of GC formation for metal-poor and metal-rich GCs could
not have been too different.  Like the color-magnitude relation, these
relations derived from globular clusters present stringent constraints
on the formation and evolution of early-type galaxies.

\end{abstract}





\section{Introduction}

Globular clusters (GCs) are found in nearly every
nearby galaxy irrespective of its luminosity or gas content.
The ubiquity and relative simplicity of these old star clusters 
makes them a fundamental tool for understanding the star
formation, metal-enrichment, and merging histories of galaxies in the
local universe.  In particular, the color distributions of GC systems
have played an important role in constraining the evolution of
elliptical galaxies.  While the color-magnitude diagram of 
early-type galaxies has garnered much interest as a strong constraint on 
the formation of ellipticals (e.g.\ Bower, Lucey, \& Ellis 1992;
Stanford, Eisenhardt, \& Dickinson 1998), 
the mean color of 
an entire galaxy is a crude tool that necessarily combines
its detailed star formation and chemical enrichment history into a
single number.  By comparison, globular clusters trace each major epoch
of star formation, and because they are individually resolved
single-age, single-metallicity systems with typical ages older than
10~Gyr, they provide a means to determine the distribution in metallicity
of the early major star forming events that built their host galaxies.

Metallicity distributions of the field star populations in
most nearby ellipticals are difficult or impossible to obtain with current
technology, therefore the color distributions of their 
GC systems (inferred to be distributions in metallicity) provide
the only direct probe of their chemical enrichment history.
With the dichotomous nature of the Milky Way's own GC system
as a starting point (Searle \& Zinn 1978), 
the availability of the Hubble Space Telescope (HST) 
over the past 15 years has enabled studies of the color
distributions of extragalactic GC systems, and has 
revealed that bimodality is a common property in
massive elliptical galaxies (Gebhardt \& Kissler-Patig 1999; Kundu \&
Whitmore 2001).  This has led to the nomenclature of blue or
``metal-poor'', and red or ``metal-rich'' GC populations, which in the
Milky Way would correspond to the halo GCs ($\langle {\rm [Fe/H]}
\rangle \sim -1.59$) and the bulge/thick disk GCs ($\langle {\rm
  [Fe/H]} \rangle \sim -0.55$) (\cote 1999).  Those dwarf elliptical galaxies
that have been studied to date exhibit purely unimodal
populations of metal-poor GCs (Lotz, Miller, \& Ferguson 2004).

Why these color distributions should be broad or bimodal---i.e.\ why there
should be distinct GC subpopulations---is a key issue
in the quest to assemble a consistent picture of bulge, halo, and 
elliptical galaxy formation (see West \etal 2004 for a review).  At
the very least, many if not all spheroidal systems cannot have 
formed in a single, isolated, monolithic starburst.  
We see in the local universe 
that massive star clusters are formed wherever there is
a high surface density of star formation (Larsen \& Richtler 2000),
and especially in major mergers (e.g.\ Zhang, Fall \& Whitmore 2001).
Thus, some proposed explanations for GC color bimodality 
invoke sequential episodes of star formation to create the
two populations, both induced by major mergers 
(Ashman \& Zepf 1992) and in isolation
(Forbes \etal 1997).  Others attempt to explain GC systems in the context of
hierarchical merging with gas dissipation in both semi-analytic 
(Beasley \etal 2002) and hydrodynamic models (Kravstov \& Gnedin 2005).
While these approaches are in some ways the most 
promising and represent the more
quantitative work applied to this problem, neither model naturally
produces bimodality.  \cote \etal (1998, 2000) showed
that multiple star forming events were {\it not} necessary to produce
bimodal metallicity distributions in a hierarchical merging framework
without gas dissipation.  We see evidence of this kind of
gas-poor merging both locally in the form of accreted dwarf
galaxies like the Sagittarius dwarf spheroidal galaxy 
(Ibata, Gilmore \& Irwin 1995), and also in mergers
of massive ellipticals in high redshift clusters (van~Dokkum \etal 1999).

It is probable that aspects of all of these scenarios (which
themselves are not mutually exclusive) are important for the formation
of globular cluster systems.  Determining the dominant
mechanisms for the formation of the GC subpopulations will come from
quantifying the detailed nature of these GCs.  Studies of
metal-poor and
metal-rich GCs in individual galaxies have already shown that they
have different spatial properties with the metal-rich GCs being more
concentrated toward the centers of galaxies (e.g.\ Geisler, Lee \& Kim
1996), and that they have
different kinematic properties (Sharples \etal 1998; Zepf \etal 2000; 
\cote \etal 2001, 2003; Peng \etal 2004).  Another approach 
to understanding GC systems, which we take in this paper, is by 
precisely quantifying their
variation as a function of host galaxy properties.
In addition, this allows us to address the
similarities and differences between GCs and field stars in a
systematic way,
providing insight into the nature of star formation and gas flows 
in the early universe.  

Previous studies have investigated GC
color or metallicity distributions as a function of host galaxy properties.  
van den Bergh (1975) and Brodie \& Huchra (1991) 
discussed the correlation between the mean
metallicity of GC systems with the mean metallicity of their host
galaxies.  It is now well-established that more luminous and metal-rich
galaxies have more metal-rich GCs.
Moreover, the mean color
of the metal-rich GC subpopulation also appears to correlate with galaxy
luminosity.  Both Forbes \etal (1997) and Kundu \& Whitmore (2001) 
found that the colors of the
metal-rich GCs correlated with host galaxy luminosity, but neither
were able to detect a correlation for the metal-poor GCs.  Forbes \&
Forte (2001) found the same to be true when the GC colors were compared
to galaxy velocity dispersion.

Larsen \etal (2001)
conducted a careful and 
homogeneous study of both GC subpopulations 
with deep HST/WFPC2 photometry in the 
F555W and F814W filters of 17 galaxies, all of which were brighter
than $M_B=-18.6$.  In the GC color distributions of these galaxies,
the mean colors of {\it both} the metal-rich and metal-poor peaks
increased with host galaxy luminosity.  They also for the first time 
found that the GC system colors had a weak correlation with host
galaxy color (expected because of the correlation with luminosity and the
color-magnitude relation of early-type galaxies).  
Of particular interest lately
is the degree to which the properties of metal-poor GCs are correlated
with their host galaxies.  The lack or weakness of any observed trends
have indicated to previous authors that the metal-poor GCs must be
``universal'' and have formed in similar sized gas fragments under
similar conditions.
Burgarella \etal (2001) compiled color data on
the blue GCs for 47 galaxies from the literature and found the trend to be
weak or insignificant.  At the other end of the galaxy luminosity function, 
Lotz \etal (2004) measured the GC color distributions of dwarf elliptical
galaxies (dEs) using HST/WFPC2 imaging of 69 dEs in the Virgo and Fornax
clusters and they too found a shallow relationship between GC color and galaxy
luminosity.  All of these studies relied on the best data at the
time, HST/WFPC2 imaging of galaxy samples with varying
degrees of heterogeneity, and almost all were based on $V$--$I$ colors.  
The installation of the Advanced Camera for
Surveys (ACS, Ford \etal 1998) now gives us the opportunity to conduct
this sort of study in a much more precise and comprehensive fashion.

The ACS Virgo Cluster Survey (ACSVCS, \cote \etal 2004, Paper I) is an
HST/ACS imaging 
program of 100 early-type galaxies in the Virgo cluster, and is designed
for studying GC systems in a deep and homogeneous manner.  The breadth
and depth of this survey makes it the most complete and
homogeneous study of extragalactic globular cluster populations ever
undertaken.  Characterizing the metallicity
distributions of GC systems is one of the key goals of this project.  
Each galaxy is imaged in the F475W and
F850LP filters ($g_{475}$ and $z_{850}$), providing twice the
wavelength baseline and metallicity sensitivity of the standard
$V$--$I$ color.  The high spatial resolution and spatial sampling of
the ACS Wide Field Camera (ACS/WFC) allows us to resolve the half-light
radii of GCs at the distance of Virgo, facilitating the selection of
GC candidates and enabling studies of the GC sizes themselves (\jordan\ \etal
2005a, Paper X).  Our images also provide high precision photometry and surface
brightness profiles of the galaxies themselves (Ferrarese \etal 2005,
Paper VI),
and their nuclei (\cote\ \etal 2005, Paper VIII),
as well as distances to each galaxy using the method of surface
brightness fluctuations (Mei \etal 2005a,b Papers IV and V).  
These data have also begun to show the connection between globular clusters and
ultracompact dwarf galaxies (Ha{\c s}egan \etal 2005, Paper VII).
Future studies (Mieske et al., in prep) will also use the ACSVCS data
to address the topic of color-magnitude relations for the GCs
themselves (e.g.\ Harris \etal 2005).  Taken together, we are
able to present a complete picture of these Virgo galaxies and their
GC systems.

\section{Data and Catalogs}
\subsection{Data Reduction}
We have reduced our ACS/WFC images using a dedicated pipeline that is
described by \jordan \etal (2004a,b, Papers II, III).  Briefly, each image is
combined and cleaned of cosmic rays using the Pyraf task 
{\it multidrizzle} (Koekemoer \etal 2002).  We then create models of the
galaxy in each filter for the purposes of subtracting the galaxy
light.  After model subtraction, we iterate with SExtractor (Bertin
1996) to mask objects, subtract any residual background, and do our
final object detection using estimates of both the image noise and the
noise present due to surface brightness fluctuations.  Objects are only
accepted if they are detected 
in both filters.  After a selection on magnitude and ellipticity
to reject obvious background galaxies, we use the program KINGPHOT
(\jordan \etal 2005a) to measure magnitudes and sizes for all candidate GCs. 
KINGPHOT finds the best-fit King model parameters for each object
given the point spread function (PSF) in that filter and at that location.
Aperture magnitudes are measured on
the best-fit PSF-convolved King model.  
For total
magnitudes, we integrate the flux of the model to the limit of the PSF
and apply a GC size-dependent aperture correction.  This correction
is determined
by convolving King models of different half-light radii with both
the Sirianni \etal
PSF (F475W), and a PSF derived from bright stars in the Galactic
globular cluster 47 Tuc (F850LP).  
For the purpose of measuring GC
colors, we measure magnitudes within a 4-pixel radius aperture and
then apply the average aperture correction for a GC with a half-light
radius of 3~pc and a concentration of 1.5.
The result is a catalog of total 
magnitudes, \gz\ colors, half-light radii ($r_h$) and concentrations for
each object.  Magnitudes and colors are corrected for foreground
extinction using the reddening maps of Schlegel, Finkbeiner \& Davis
(1998), and extinction ratios for the spectral energy distribution of
a G2 star (Paper II).  In this paper, we use $g$--$z$ to mean
$g_{475}$--$z_{850}$ unless explicitly stated otherwise.
 
\begin{deluxetable}{ccc}
\tablewidth{0pt}
\tablecaption{Control Fields\label{table:controls}}
\tablehead{
\colhead{RA(J2000)} &
\colhead{Dec(J2000)} &
\colhead{HST Program ID}
}
\startdata
02:06:33.85 & $+20$:53:13.36 & 9488 \\
02:17:20.72 & $-04$:44:41.01 & 9488 \\
04:40:44.48 & $-45$:44:21.24 & 9488 \\
09:54:35.32 & $+69$:49:10.54 & 9575 \\
12:10:53.82 & $+39$:14:06.33 & 9488 \\
12:10:33.95 & $+39$:35:22.63 & 9575 \\
12:25:51.14 & $+00$:04:08.60 & 9488 \\
12:38:32.53 & $+62$:26:54.43 & 9488 \\
12:43:30.12 & $+11$:49:20.62 & 9488 \\
13:41:40.42 & $+28$:33:37.39 & 9488 \\
13:53:57.73 & $+69$:28:26.75 & 9488 \\
15:16:15.80 & $+07$:09:43.46 & 9488 \\
15:43:32.26 & $+59$:21:37.66 & 9488 \\
15:44:50.76 & $+59$:02:07.37 & 9488 \\
22:12:37.23 & $-83$:54:18.95 & 9575 \\
22:22:57.51 & $-72$:23:21.07 & 9488 \\
23:46:49.49 & $+12$:44:53.35 & 9488 \\
\enddata
\end{deluxetable}


\bigskip
\bigskip
\subsection{Control Fields}
One of the main problems with GC studies, particularly for fainter
galaxies, is the contamination from background galaxies.  For dwarf
galaxies where there may only be a few GCs, the contamination from the
background is often larger than the signal.  For brighter
galaxies, the light from the galaxy itself creates a spatially varying
detection efficiency that complicates the comparison to blank fields.
This is a problem that requires careful treatment in order to minimize
systematic uncertainties, especially in low-luminosity systems.

In order to address this issue, we use archival ACS/WFC
imaging of 17 blank, high-latitude fields that 
have been observed with the F475W and F850LP filters to the same or
greater depth than our ACSVCS observations.  These
images were taken as part of two ACS Pure Parallel programs
(GO-9488,GO-9575), and
are well-suited to the purpose of understanding the background and
foreground contamination in our GC samples
(Table~\ref{table:controls}).  Each one of these images
was reduced with our pipeline in exactly the same way as our program
data.  Since there were no large foreground galaxies in any of the control
fields, and the exposure times were longer than for our ACSVCS
images, all the control data go deeper than our program data.
Hence for each galaxy, we created a ``custom'' control sample by
redoing the detection {\it as if} there was a Virgo galaxy in front of
of each blank field, and each field had the same exposure time as our
ACSVCS fields.  To do this, we used the noise images for our program
fields that are generated in
the data reduction.  Thus, for each
control field we can produce the catalog of objects that we would have
detected had the exposure times matched our ACSVCS program images, and
had there been a particular ACSVCS galaxy in the foreground.  

We find that using the control fields is consistent with and 
superior to using the local background.  The degree of
field-to-field cosmic variance for GC candidates in our control fields
is on the order of the Poisson noise.  Given our efficient selection of
GC-like objects, there are not many background objects that make the
cut for GC selection (see below).  
For example, in our faintest galaxy, VCC~1661, we
only expect on average about $\sim9 \pm 4$ contaminants to the sample
of GC candidates.  Determining this
background locally as opposed to over many control fields, while
possibly more representative of the true background, introduces a
large error due to counting noise of small numbers.  In general, we find
that the number of contaminants that we estimate from the control fields
is a good match to the number of contaminants found in the dwarf galaxy
fields, where the number of objects is dominated by the background
(see Peng \etal 2005).

\subsection{Globular Cluster Selection}
The selection of a clean and complete catalog of GCs is an essential but
difficult part of all GC studies.  The ability to resolve and measure
sizes for GCs in Virgo with HST provides much greater leverage to separate
GCs from foreground stars and background galaxies.  In addition, our
custom control fields allow us to isolate the location of our
contaminating population in the multidimensional parameter space.  This
efficient selection allows us to avoid stringent cuts on galactocentric
radius that may strongly bias measured properties of the GC system.
This approach to GC selection and background estimation makes our
study particularly unique and important for the GC systems of faint
galaxies where the number of GCs is small compared to the number of
background galaxies.  Previous studies have all required a strict selection
on galactocentric radius.

\begin{figure}
\plotone{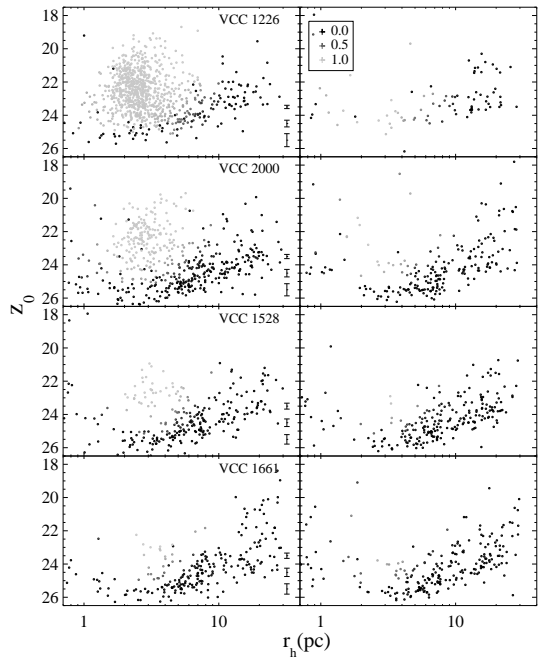}
\caption{Globular cluster selection diagrams for four galaxies
  spanning the magnitude range of our survey.  From top to bottom:
  VCC 1226 (M49, $B=9.31$), VCC 2000 ($B=11.94$), VCC 1528 ($B=14.51$),
  and VCC 1661 ($B=15.97$).  Left panels
  show the objects detected in the galaxy images.  Right panels show
  a random selection of $1/17$ of the objects detected in 17 blank sky
  control fields that have been customized for the depth reached in
  each galaxy.  Notice how the control field for VCC 1226 is shallower
  and has fewer objects because the galaxy is brighter and fills
  the frame.  Object colors are coded by the probability that they are
  globular clusters.  Error bars represent the median
  photometric error at those magnitudes. \label{fig:showselection}}
\end{figure}

While the details of the GC selection will be described in greater
detail in another paper (\jordan \etal 2005b), here we briefly outline the
technique.  We choose a broad cut on color, selecting only objects with
$0.5 < g$--$z < 2.0$, which generously spans the age and metallicity
ranges typical of old star clusters.
The strength of our selection lies in the size-magnitude diagram
(Figure~\ref{fig:showselection}), where background galaxies are
typically fainter and more extended than GCs.  This figure shows the
data and custom control fields for two galaxies in our sample.  The
custom control fields have been randomly sampled so that only $1/17$ of
the objects are plotted, equivalent to what is expected for a single
field.  It is already well known that GCs typically have a Gaussian-like
luminosity function (Harris 1991), and that the Galactic GCs have a median size
of $r_h\sim3$~pc.  This is evident in the left hand panels of
Figure~\ref{fig:showselection} as a population mostly separate from the
locus of background contaminants.  Notice also how the custom control
field includes fewer objects and is shallower for the brighter galaxy.

We apply a Gaussian kernel to the control data to produce a
nonparametric two dimensional density distribution of contaminants.  We then
fit a parametric model to the GC locus. In magnitude, we assume
a Gaussian distribution, and in size, we use a nonparametric kernel
estimate plus a power law tail, with the two joined at $r_h=7$~pc.
This model is then fit to
the data using maximum likelihood estimation --- the fitting procedure
allows us to account for variations in the GCLF turnover magnitude and
differences in the size distributions from galaxy to galaxy.  
Finally, we 
use these two density distributions to evaluate the probability that a
given object is a GC.  Typically, 
the lines of constant probability run diagonally from the faint, compact
region of the diagram to the bright, extended region (seen in
Figure~\ref{fig:showselection}).  However, even at a constant probability
cut at a value of 0.5, this dividing line is not exactly the same from
galaxy to galaxy.  In galaxies with more GCs, it is more likely that
objects in the ``contaminant'' locus are actually GCs, and hence they
will be assigned a higher probability.  For the purposes of this paper,
all of our GC catalogs will use objects with GC probabilities greater
than 0.5.

Recently, some investigations have revealed the
existence of faint, extended star clusters in NGC~1023, NGC~3384 (Larsen
\& Brodie 2000; Brodie \& Larsen 2002), and other nearby spiral
(Chandar, Whitmore, \& Lee 2004) and dwarf galaxies (Sharina, Puzia, 
\& Marakov 2005).  Because these clusters
typically have similar sizes and magnitudes to background galaxies and
their nature is still uncertain, our selection cuts do not include them
in our GC sample and we do not include them in our analysis.  Faint,
extended star clusters in the ACSVCS galaxies will be treated in
a separate paper (Peng \etal 2005).

\section{Results}
\subsection{Color Distributions of Globular Cluster Systems}

\begin{figure*}
\plotone{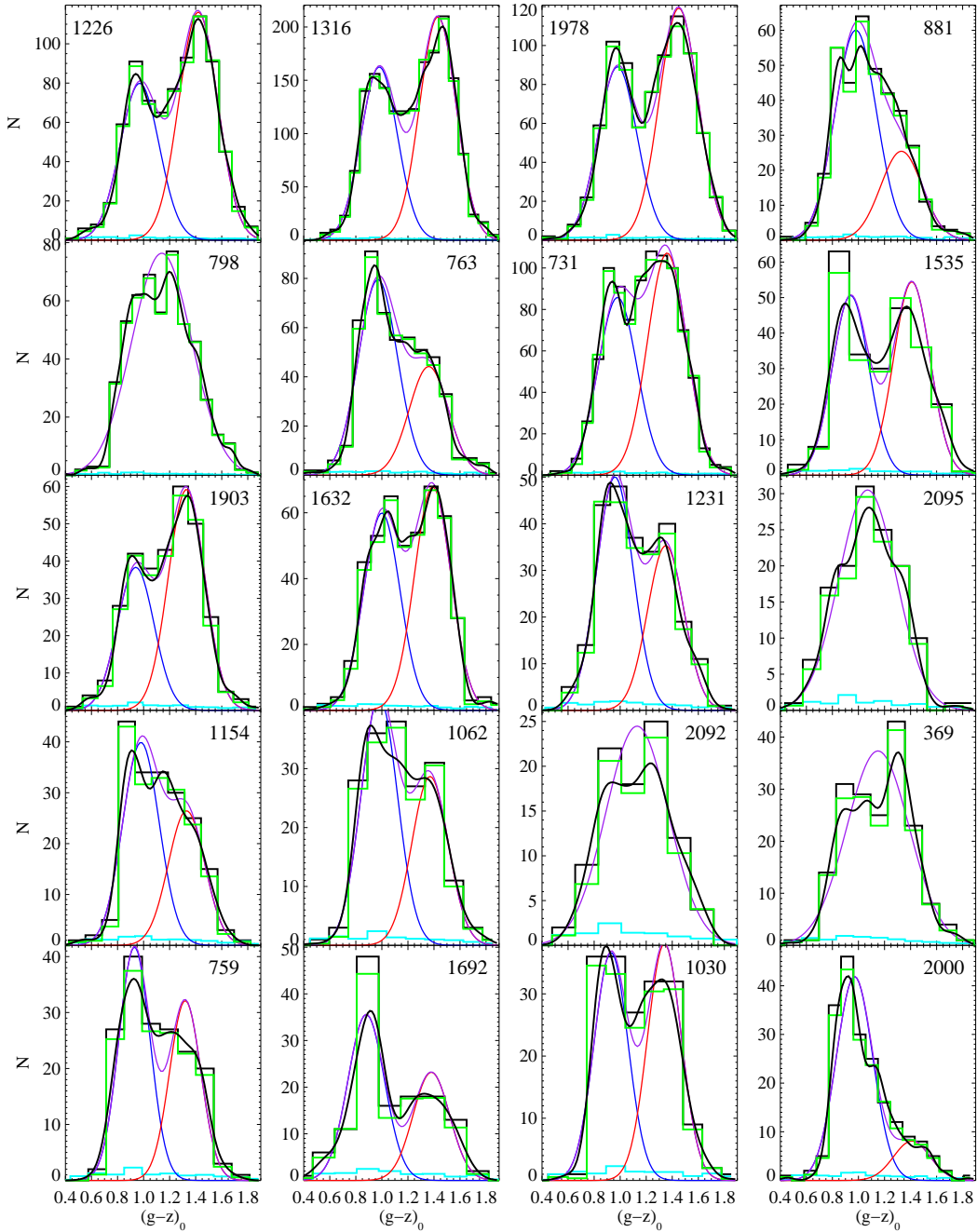}
\caption{Globular cluster $(g$--$z)_0$ distributions
 for all 100 ACSVCS galaxies.  
 We plot color histograms of GC candidates (black),
 expected contaminants (cyan), and statistically cleaned GC
 distributions (green).  The black curve represent a nonparametric kernel
 density estimate of the cleaned distribution.  In cases where the
 distribution is likely bimodal, we plot the red and blue Gaussian
 components as determined by the KMM estimates, as well as their sum
 (purple).  In the unimodal cases, we plot the best fit single Gaussian
 to the entire distribution (purple).\label{fig:colorhists}}
\end{figure*}

\begin{figure*}
\figurenum{2}
\plotone{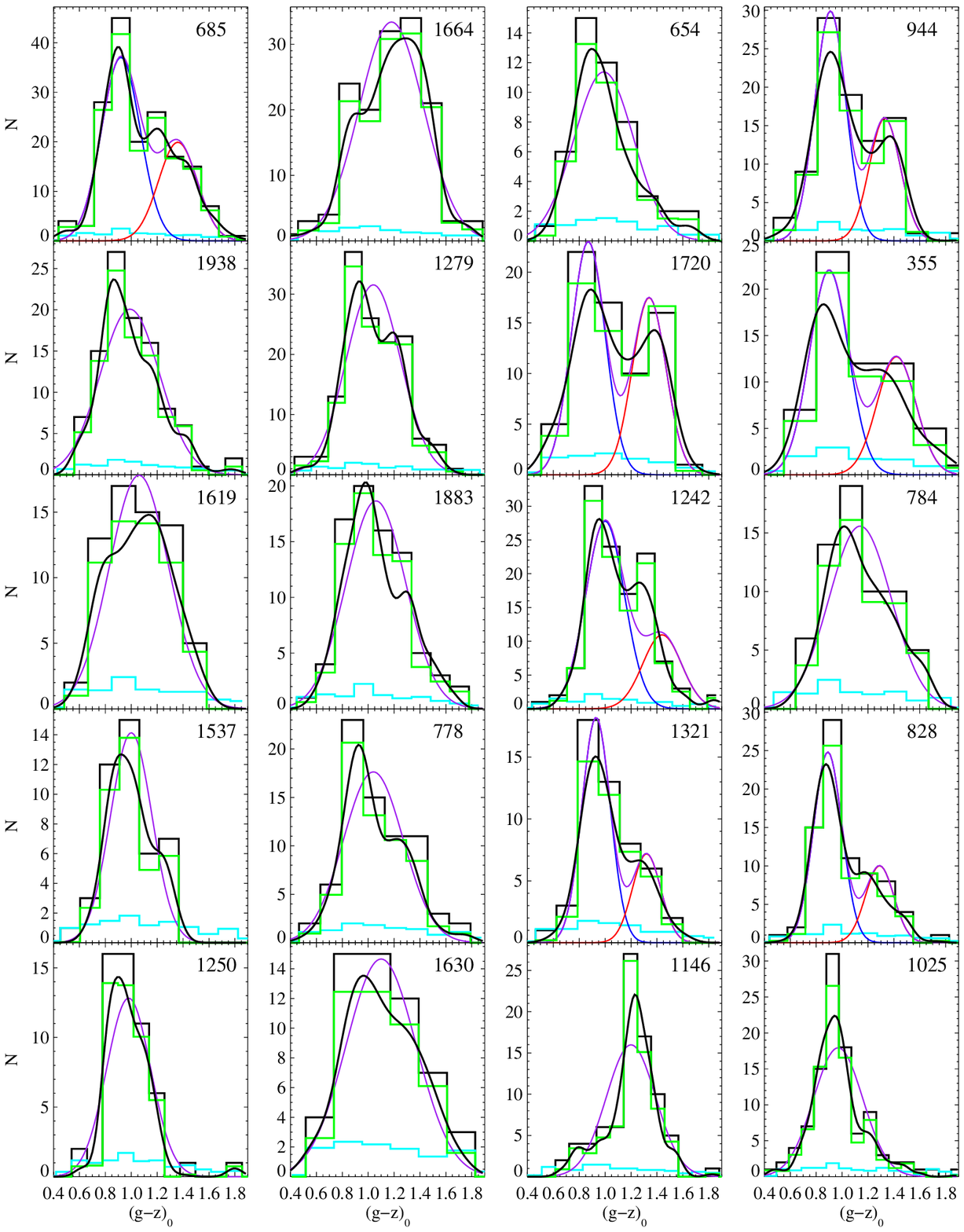}
\caption{continued.  Globular cluster $(g$--$z)_0$ distributions.}
\end{figure*}

\begin{figure*}
\figurenum{2}
\plotone{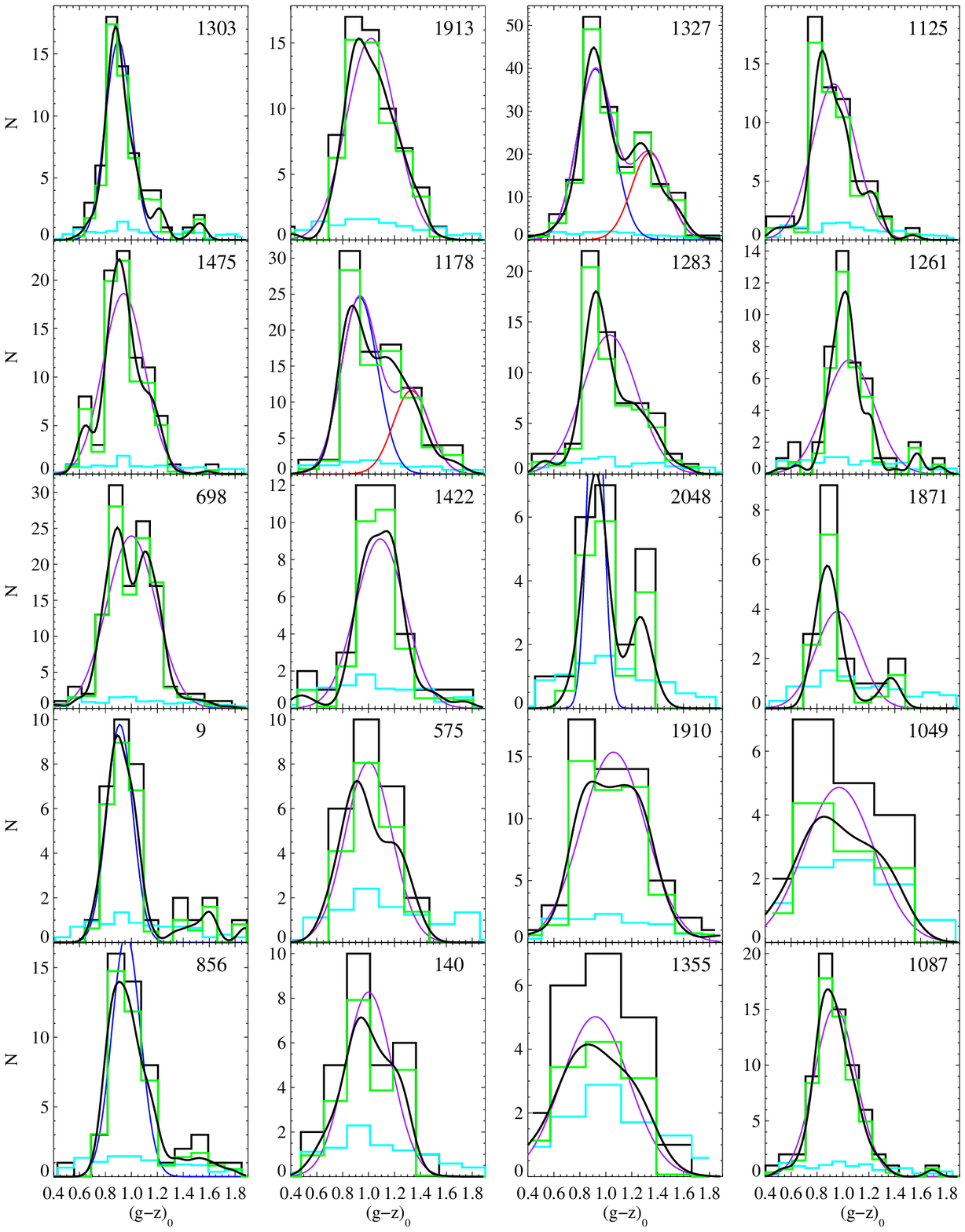}
\caption{continued.  Globular cluster $(g$--$z)_0$ distributions.}
\end{figure*}

\begin{figure*}
\figurenum{2}
\plotone{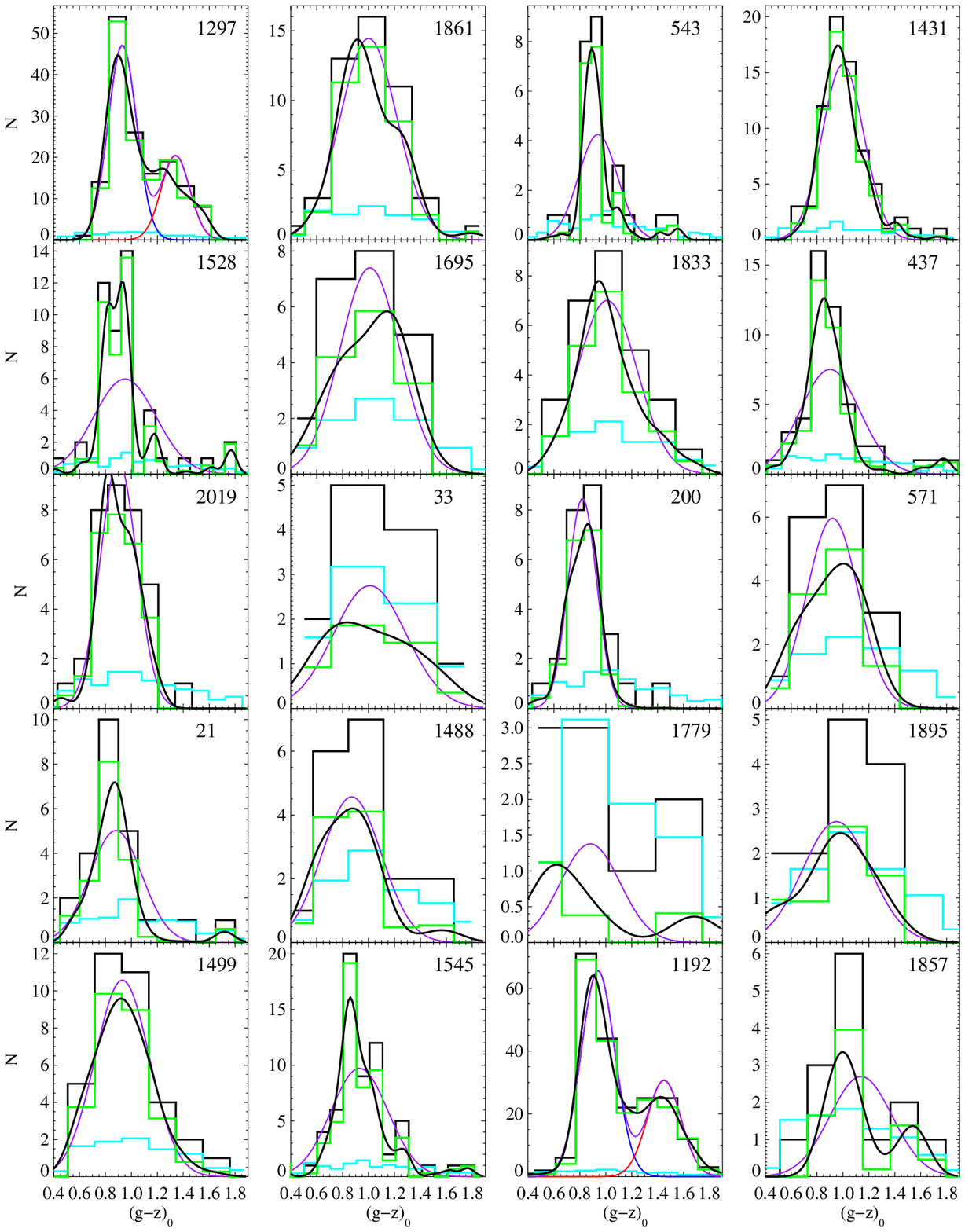}
\caption{continued.  Globular cluster $(g$--$z)_0$ distributions.}
\end{figure*}

\begin{figure*}
\figurenum{2}
\plotone{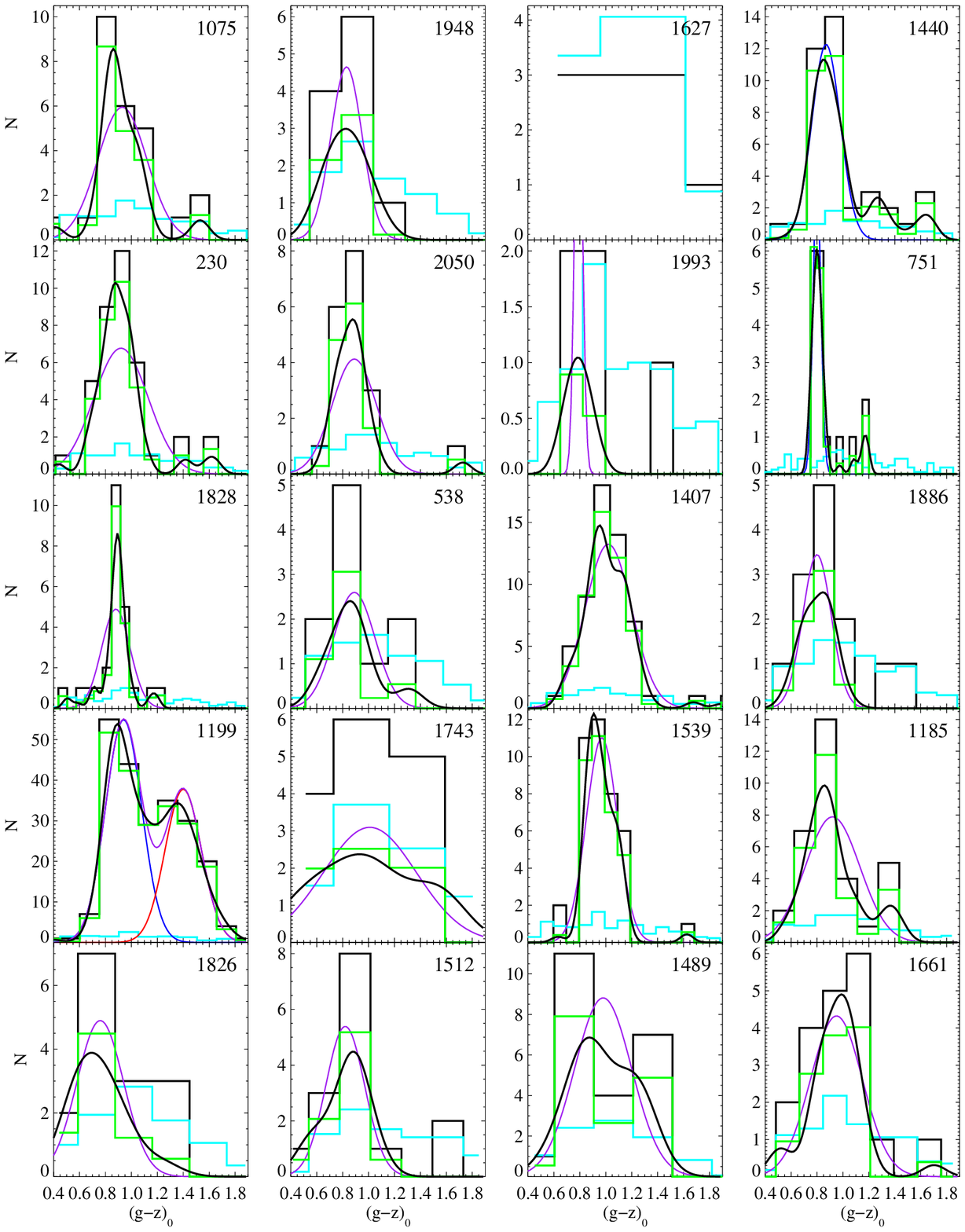}
\caption{continued.  Globular cluster $(g$--$z)_0$ distributions.}
\end{figure*}

In Figure~\ref{fig:colorhists}, we show the $(g$--$z)_0$ color
distributions for  
the GC systems of 100 ACSVCS galaxies, ordered by decreasing total
$B$ magnitude of the host galaxy.  The histograms were created by
binning the data with an ``optimal'' bin size, which is related to the
width of the distribution and the total number of objects as
$2 IQR n^{-1/3}$, where $IQR$ is the interquartile range (the color range
that includes the second and third quartile of the ranked color distribution),
and $n$ is the total number of objects (Izenman 1991).  The bin size is
not allowed to 
be smaller than the mean photometric
error.  The histograms of the GC data are shown in black, and histograms
of the expected contamination as measured from the custom control
fields are shown in cyan.  

While the level of
contamination is negligible for the brighter galaxies, the background is
a significant problem for the fainter galaxies.  We create
statistically cleaned samples of GCs by using a Monte Carlo procedure.  
For each GC, we calculate the probability that it is a contaminant by
using a nonparametric density estimate of the control data as
compared to the program galaxy data at that color.  Based on this
probability, we randomly include or do not include this object 
(with replacement) from our generated sample.  Iterating 100 times for each
galaxy, we can produce an average color distribution that is cleaned of
contaminants.  These histograms are plotted in green.  
Kernel density estimates of the cleaned distribution are
overplotted as black curves.

The cleaned and smoothed color distributions show
obvious bimodality for most of the brighter galaxies, something which is
expected from previous investigations (e.g.\ Gebhardt \& Kissler-Patig
1998; Larsen \etal 2001, Kundu \& Whitmore 2001).  For intermediate
luminosities and fainter 
galaxies, the situation becomes less obvious as the number of GCs per
galaxy decreases.  Yet, it is clear that the fraction of red GCs
decreases as the galaxy itself becomes fainter.

The properties of the GC color distribution are often quantified using
the Kaye's Mixture Model (KMM; McLachlan \& Basford 1988; Ashman,
Bird, \& Zepf 1994) to fit two Gaussians to the data using the
expectation-maximization (EM) method.  We fit the homoscedastic
case of this model, where $\sigma$ is the same for both Gaussians.
Constraining $\sigma$ reduces the number of free parameters and gain
leverage on noisy data in which there are generally large errors when
fitting for $\sigma$ individually.  We apply KMM to the
each of the probabilistically cleaned GC color samples to determine
the means of the blue and red peaks in the color distribution.  We
determine that the two Gaussian model is a better fit to the data than the
one Gaussian model if the ``p-value'' for the bimodal model is less
than 0.05.  For the cases where bimodality is deemed significant 
($p\leq 0.05$),
Figure~\ref{fig:colorhists} shows the two Gaussians and their sum.
For the unimodal case, the single best-fit Gaussian is plotted.  In
some galaxies, the distribution was determined to be bimodal, but the
number of GCs in the red peak was not significant over the background at
greater than 2$\sigma$.  In these cases, we reject the bimodal
hypothesis because the red objects are likely to be background galaxies.  The
results for KMM, and the parameters for the best fit one or two
Gaussian models are presented in Table~\ref{table:kmmtable}.  This
table includes the means of the blue and red peaks, the common sigma,
the fraction of GCs that are determined to be in the red subcomponent,
the total number of GCs (after accounting for the background), the
p-value of the two Gaussian hypothesis and its associated error, and
the mean and sigma for the entire distribution.  In cases where the
red peak was less than 2$\sigma$ above the background, only the mean
color of the blue peak is reported.  In cases where the total number
of GCs was statistically equal to zero (see next section), no values
could be reliably estimated.

\subsection{Notes on Special Galaxies}
A few galaxies are noteworthy in that they appear somewhat different
from others of similar luminosity.  For a more in-depth description of
every galaxy in the sample, see Ferrarese \etal (2005, Paper VI).
Four galaxies in particular, despite their low luminosities, have prominent
blue and red peaks and have large numbers of GC.  This is due to their 
proximity to the two giants of the cluster, M87 and M49, and we are
likely observing the GC systems of their larger
neighbors.  Interestingly, all four are also significantly redder than one
would expect for galaxies of their luminosity, and both the number and
mean color of the red GCs in these galaxies is similar to those seen
in galaxies of the same $(g$--$z)$ color.  This suggests that they may be
remnants of what were once larger, more luminous systems.

\noindent{\bf VCC1327/NGC4486A}: This dwarf elliptical is only
7.5\arcmin\ away from M87 (NGC~4486), and its GCs are likely to be
dominated by those of the nearby cD galaxy.

\noindent {\bf VCC 1297/NGC4486B}:  This compact elliptical galaxy is similar
to M32 in appearance, and is only 7.3\arcmin\ away from M87.  
Some or all of its GCs may in fact belong to the M87 GC
system, and its current size and luminosity may be the result of
significant tidal stripping. 

\noindent{\bf VCC 1192/NGC4467}:  This elliptical galaxy is also compact
in appearance and is only 4.2\arcmin\ away from M49 (NGC~4472).  Its GC
system may be dominated by that of M49.

\noindent{\bf VCC 1199}:  This elliptical is also only 4.5\arcmin\ from
M49.  Like the other small galaxies near giant ellipticals, it
displays a prominent red peak despite its low luminosity.

\noindent Other galaxies of note are:

\noindent{\bf VCC 1146/NGC4458}:  This galaxy is unique in the sample
in that its GC system is dominated by a single red peak of GCs. 
The red GC fraction of 0.84 estimated by KMM is not only much larger than 
the value of 0.3 expected for a galaxy
of this luminosity, but is much higher than the 0.6 fraction seen in
the giant ellipticals.

\noindent{\bf VCC 798}: This galaxy appears to be the best candidate
in our sample for having a trimodal color distribution.

\noindent{\bf VCC 731}: Appears to have significantly more GCs than
other galaxies of comparable luminosity.  The excess appears to be
due to an large number of red GCs.

\noindent{\bf VCC 1692}:  The blue and red GCs are particularly well
separated in color.

{\centering
\begin{deluxetable*}{crrrrrrrrrrrrrrrr}
\tablewidth{0pt}
\tabletypesize{\tiny}
\tablecaption{Color Distributions: Bimodal and Unimodal Parameters \label{table:kmmtable}}
\tablehead{
\colhead{No.} & 
\colhead{VCC} & 
\colhead{$\mu_{blue}$} & \colhead{$\mu_{blue,err}$} & 
\colhead{$\mu_{red}$} & \colhead{$\mu_{red,err}$} & 
\colhead{$\sigma$} & \colhead{$\sigma_{err}$} & 
\colhead{$f_{red}$} & \colhead{$f_{red,err}$} & 
\colhead{$N_{GC}$} & 
\colhead{$p$} & \colhead{$p_{err}$} & 
\colhead{$\mu_1$} & \colhead{$\mu_{1,err}$} & 
\colhead{$\sigma_1$} & \colhead{$\sigma_{1,err}$} \\ 
\colhead{(1)} &
\colhead{(2)} &
\colhead{(3)} &
\colhead{(4)} &
\colhead{(5)} &
\colhead{(6)} &
\colhead{(7)} &
\colhead{(8)} &
\colhead{(9)} &
\colhead{(10)} &
\colhead{(11)} &
\colhead{(12)} &
\colhead{(13)} &
\colhead{(14)} &
\colhead{(15)} &
\colhead{(16)} &
\colhead{(17)}
}
\startdata
   1 & 1226 &  0.97 &  0.01 &  1.42 &  0.01 &  0.15 &  0.01 &  0.59 &  0.03 &  749 & $< 0.01$ & $< 0.01$ &  1.24 &  0.01 &  0.27 &  0.01 \\
   2 & 1316 &  0.98 &  0.01 &  1.43 &  0.01 &  0.14 &  0.01 &  0.56 &  0.02 & 1723 & $< 0.01$ & $< 0.01$ &  1.23 &  0.01 &  0.26 &  0.00 \\
   3 & 1978 &  0.98 &  0.01 &  1.45 &  0.01 &  0.15 &  0.01 &  0.57 &  0.03 &  791 & $< 0.01$ & $< 0.01$ &  1.25 &  0.01 &  0.27 &  0.00 \\
   4 &  881 &  0.98 &  0.03 &  1.33 &  0.03 &  0.16 &  0.01 &  0.30 &  0.10 &  353 &  0.01 &  0.02 &  1.09 &  0.01 &  0.23 &  0.01 \\
   5 &  798 &  1.02 &  0.02 &  1.34 &  0.03 &  0.17 &  0.01 &  0.37 &  0.09 &  503 &  0.07 &  0.15 &  1.14 &  0.01 &  0.23 &  0.01 \\
   6 &  763 &  0.97 &  0.01 &  1.36 &  0.01 &  0.15 &  0.01 &  0.36 &  0.04 &  489 & $< 0.01$ & $< 0.01$ &  1.11 &  0.01 &  0.24 &  0.01 \\
   7 &  731 &  0.98 &  0.01 &  1.36 &  0.01 &  0.15 &  0.01 &  0.56 &  0.02 &  889 & $< 0.01$ & $< 0.01$ &  1.19 &  0.01 &  0.24 &  0.00 \\
   8 & 1535 &  0.94 &  0.01 &  1.41 &  0.01 &  0.14 &  0.01 &  0.52 &  0.04 &  234 & $< 0.01$ & $< 0.01$ &  1.18 &  0.02 &  0.28 &  0.01 \\
   9 & 1903 &  0.94 &  0.02 &  1.33 &  0.01 &  0.14 &  0.01 &  0.61 &  0.03 &  296 & $< 0.01$ &  0.01 &  1.18 &  0.01 &  0.24 &  0.01 \\
  10 & 1632 &  1.00 &  0.01 &  1.39 &  0.01 &  0.14 &  0.01 &  0.53 &  0.03 &  437 & $< 0.01$ & $< 0.01$ &  1.21 &  0.01 &  0.24 &  0.01 \\
  11 & 1231 &  0.96 &  0.02 &  1.35 &  0.01 &  0.14 &  0.01 &  0.41 &  0.03 &  240 & $< 0.01$ & $< 0.01$ &  1.12 &  0.01 &  0.24 &  0.01 \\
  12 & 2095 &  0.92 &  0.05 &  1.22 &  0.04 &  0.16 &  0.02 &  0.53 &  0.16 &  123 &  0.43 &  0.33 &  1.07 &  0.02 &  0.22 &  0.01 \\
  13 & 1154 &  0.98 &  0.02 &  1.33 &  0.03 &  0.14 &  0.01 &  0.40 &  0.09 &  184 &  0.01 &  0.05 &  1.12 &  0.01 &  0.22 &  0.01 \\
  14 & 1062 &  0.98 &  0.02 &  1.37 &  0.02 &  0.14 &  0.01 &  0.40 &  0.07 &  169 & $< 0.01$ &  0.01 &  1.14 &  0.02 &  0.24 &  0.01 \\
  15 & 2092 &  0.94 &  0.04 &  1.33 &  0.04 &  0.15 &  0.02 &  0.50 &  0.08 &   83 &  0.11 &  0.14 &  1.13 &  0.03 &  0.24 &  0.02 \\
  16 &  369 &  0.95 &  0.02 &  1.32 &  0.02 &  0.15 &  0.02 &  0.54 &  0.07 &  170 &  0.07 &  0.13 &  1.15 &  0.02 &  0.24 &  0.01 \\
  17 &  759 &  0.93 &  0.02 &  1.32 &  0.02 &  0.12 &  0.01 &  0.44 &  0.07 &  161 & $< 0.01$ & $< 0.01$ &  1.10 &  0.01 &  0.23 &  0.01 \\
  18 & 1692 &  0.88 &  0.01 &  1.38 &  0.03 &  0.14 &  0.01 &  0.39 &  0.05 &  122 & $< 0.01$ & $< 0.01$ &  1.08 &  0.03 &  0.28 &  0.01 \\
  19 & 1030 &  0.93 &  0.02 &  1.34 &  0.02 &  0.13 &  0.01 &  0.51 &  0.06 &  165 & $< 0.01$ & $< 0.01$ &  1.14 &  0.02 &  0.24 &  0.01 \\
  20 & 2000 &  0.97 &  0.02 &  1.41 &  0.04 &  0.14 &  0.01 &  0.16 &  0.05 &  186 & $< 0.01$ & $< 0.01$ &  1.05 &  0.02 &  0.22 &  0.01 \\
  21 &  685 &  0.92 &  0.02 &  1.36 &  0.04 &  0.15 &  0.02 &  0.35 &  0.08 &  156 & $< 0.01$ &  0.01 &  1.07 &  0.03 &  0.26 &  0.02 \\
  22 & 1664 &  0.91 &  0.10 &  1.31 &  0.04 &  0.18 &  0.03 &  0.69 &  0.15 &  132 &  0.17 &  0.23 &  1.18 &  0.02 &  0.25 &  0.02 \\
  23 &  654 &  0.92 &  0.03 & \nodata & \nodata &  0.15 &  0.02 & \nodata & \nodata &   42 &  0.06 &  0.09 &  0.99 &  0.03 &  0.23 &  0.02 \\
  24 &  944 &  0.91 &  0.02 &  1.33 &  0.03 &  0.12 &  0.01 &  0.35 &  0.07 &   81 &  0.01 &  0.03 &  1.06 &  0.02 &  0.24 &  0.01 \\
  25 & 1938 &  0.92 &  0.04 &  1.36 &  0.14 &  0.16 &  0.02 &  0.17 &  0.09 &   89 &  0.07 &  0.15 &  0.99 &  0.03 &  0.23 &  0.02 \\
  26 & 1279 &  0.93 &  0.02 &  1.25 &  0.04 &  0.14 &  0.02 &  0.35 &  0.12 &  128 &  0.21 &  0.29 &  1.04 &  0.02 &  0.21 &  0.01 \\
  27 & 1720 &  0.87 &  0.02 &  1.34 &  0.03 &  0.13 &  0.01 &  0.43 &  0.07 &   64 & $< 0.01$ &  0.01 &  1.08 &  0.04 &  0.27 &  0.02 \\
  28 &  355 &  0.90 &  0.04 &  1.42 &  0.06 &  0.15 &  0.02 &  0.37 &  0.08 &   52 &  0.02 &  0.04 &  1.09 &  0.04 &  0.29 &  0.02 \\
  29 & 1619 &  0.87 &  0.04 &  1.23 &  0.05 &  0.14 &  0.01 &  0.55 &  0.09 &   56 &  0.19 &  0.21 &  1.06 &  0.03 &  0.23 &  0.02 \\
  30 & 1883 &  0.95 &  0.03 &  1.35 &  0.03 &  0.14 &  0.02 &  0.27 &  0.09 &   73 &  0.07 &  0.13 &  1.06 &  0.03 &  0.23 &  0.02 \\
  31 & 1242 &  1.00 &  0.05 &  1.44 &  0.17 &  0.15 &  0.03 &  0.28 &  0.16 &  108 &  0.03 &  0.06 &  1.11 &  0.02 &  0.22 &  0.02 \\
  32 &  784 &  1.01 &  0.03 &  1.41 &  0.06 &  0.15 &  0.02 &  0.31 &  0.10 &   55 &  0.15 &  0.17 &  1.14 &  0.03 &  0.24 &  0.02 \\
  33 & 1537 &  0.91 &  0.02 &  1.20 &  0.04 &  0.09 &  0.01 &  0.28 &  0.11 &   37 &  0.17 &  0.22 &  1.00 &  0.02 &  0.16 &  0.01 \\
  34 &  778 &  0.93 &  0.05 &  1.37 &  0.12 &  0.14 &  0.03 &  0.27 &  0.13 &   61 &  0.09 &  0.14 &  1.04 &  0.03 &  0.23 &  0.02 \\
  35 & 1321 &  0.93 &  0.03 &  1.32 &  0.04 &  0.11 &  0.02 &  0.28 &  0.07 &   43 &  0.02 &  0.04 &  1.04 &  0.02 &  0.21 &  0.02 \\
  36 &  828 &  0.89 &  0.02 &  1.29 &  0.04 &  0.11 &  0.01 &  0.29 &  0.05 &   69 &  0.01 &  0.02 &  1.00 &  0.02 &  0.21 &  0.01 \\
  37 & 1250 &  0.90 &  0.04 & \nodata & \nodata &  0.09 &  0.03 & \nodata & \nodata &   46 &  0.13 &  0.30 &  0.98 &  0.02 &  0.17 &  0.04 \\
  38 & 1630 &  0.92 &  0.05 &  1.37 &  0.07 &  0.16 &  0.02 &  0.41 &  0.17 &   46 &  0.17 &  0.21 &  1.10 &  0.03 &  0.27 &  0.02 \\
  39 & 1146 &  0.90 &  0.11 &  1.28 &  0.12 &  0.13 &  0.02 &  0.81 &  0.19 &   72 &  0.07 &  0.22 &  1.20 &  0.02 &  0.19 &  0.02 \\
  40 & 1025 &  0.90 &  0.08 &  1.31 &  0.15 &  0.14 &  0.02 &  0.16 &  0.19 &   89 &  0.10 &  0.25 &  0.97 &  0.02 &  0.19 &  0.02 \\
  41 & 1303 &  0.90 &  0.02 & \nodata & \nodata &  0.10 &  0.01 & \nodata & \nodata &   53 & $< 0.01$ &  0.01 &  0.94 &  0.02 &  0.17 &  0.03 \\
  42 & 1913 &  0.92 &  0.11 &  1.19 &  0.09 &  0.14 &  0.03 &  0.36 &  0.21 &   56 &  0.45 &  0.40 &  1.02 &  0.03 &  0.19 &  0.02 \\
  43 & 1327 &  0.92 &  0.02 &  1.34 &  0.03 &  0.14 &  0.01 &  0.34 &  0.05 &  161 & $< 0.01$ & $< 0.01$ &  1.06 &  0.02 &  0.24 &  0.01 \\
  44 & 1125 &  0.88 &  0.02 &  1.19 &  0.14 &  0.13 &  0.02 &  0.16 &  0.16 &   53 &  0.34 &  0.36 &  0.93 &  0.02 &  0.17 &  0.02 \\
  45 & 1475 &  0.86 &  0.08 &  1.19 &  0.22 &  0.12 &  0.02 &  0.27 &  0.27 &   76 &  0.33 &  0.34 &  0.94 &  0.02 &  0.16 &  0.02 \\
  46 & 1178 &  0.93 &  0.04 &  1.33 &  0.08 &  0.14 &  0.02 &  0.32 &  0.12 &   80 &  0.02 &  0.03 &  1.06 &  0.03 &  0.23 &  0.02 \\
  47 & 1283 &  0.94 &  0.03 &  1.29 &  0.08 &  0.13 &  0.03 &  0.25 &  0.09 &   56 &  0.21 &  0.30 &  1.03 &  0.03 &  0.21 &  0.03 \\
  48 & 1261 &  0.99 &  0.09 & \nodata & \nodata &  0.13 &  0.03 & \nodata & \nodata &   37 &  0.21 &  0.41 &  1.05 &  0.03 &  0.19 &  0.04 \\
  49 &  698 &  0.90 &  0.15 &  1.33 &  0.24 &  0.15 &  0.03 &  0.30 &  0.31 &  108 &  0.14 &  0.25 &  1.00 &  0.02 &  0.19 &  0.02 \\
  50 & 1422 &  0.87 &  0.26 &  1.21 &  0.19 &  0.15 &  0.05 &  0.51 &  0.40 &   29 &  0.38 &  0.44 &  1.09 &  0.04 &  0.19 &  0.04 \\
  51 & 2048 &  0.92 &  0.02 & \nodata & \nodata &  0.06 &  0.01 & \nodata & \nodata &   16 &  0.02 &  0.06 &  1.01 &  0.04 &  0.17 &  0.03 \\
  52 & 1871 &  0.88 &  0.02 & \nodata & \nodata &  0.06 &  0.01 & \nodata & \nodata &   13 &  0.08 &  0.22 &  0.96 &  0.05 &  0.17 &  0.05 \\
  53 &    9 &  0.91 &  0.03 & \nodata & \nodata &  0.10 &  0.03 & \nodata & \nodata &   27 &  0.01 &  0.07 &  1.01 &  0.06 &  0.25 &  0.07 \\
  54 &  575 &  0.88 &  0.03 &  1.21 &  0.05 &  0.09 &  0.02 &  0.36 &  0.14 &   19 &  0.14 &  0.20 &  1.00 &  0.05 &  0.18 &  0.02 \\
  55 & 1910 &  0.92 &  0.15 &  1.51 &  0.30 &  0.18 &  0.05 &  0.29 &  0.30 &   47 &  0.15 &  0.19 &  1.06 &  0.03 &  0.25 &  0.03 \\
  56 & 1049 &  0.75 &  0.11 & \nodata & \nodata &  0.13 &  0.04 & \nodata & \nodata &   11 &  0.36 &  0.32 &  0.97 &  0.08 &  0.27 &  0.05 \\
  57 &  856 &  0.96 &  0.03 & \nodata & \nodata &  0.11 &  0.01 & \nodata & \nodata &   42 & $< 0.01$ &  0.01 &  1.02 &  0.03 &  0.21 &  0.04 \\
  58 &  140 &  0.83 &  0.10 &  1.15 &  0.08 &  0.11 &  0.02 &  0.50 &  0.25 &   21 &  0.35 &  0.26 &  1.00 &  0.04 &  0.18 &  0.03 \\
  59 & 1355 &  0.69 &  0.13 & \nodata & \nodata &  0.14 &  0.04 & \nodata & \nodata &   12 &  0.41 &  0.29 &  0.92 &  0.07 &  0.26 &  0.05 \\
  60 & 1087 &  0.87 &  0.11 &  1.29 &  0.32 &  0.13 &  0.02 &  0.24 &  0.30 &   59 &  0.31 &  0.40 &  0.94 &  0.02 &  0.16 &  0.03 \\
  61 & 1297 &  0.93 &  0.01 &  1.34 &  0.03 &  0.11 &  0.01 &  0.30 &  0.04 &  142 & $< 0.01$ & $< 0.01$ &  1.05 &  0.02 &  0.22 &  0.01 \\
  62 & 1861 &  0.91 &  0.05 &  1.35 &  0.22 &  0.15 &  0.03 &  0.25 &  0.12 &   39 &  0.27 &  0.32 &  1.00 &  0.04 &  0.22 &  0.03 \\
  63 &  543 &  0.89 &  0.07 & \nodata & \nodata &  0.07 &  0.03 & \nodata & \nodata &   19 &  0.08 &  0.21 &  0.94 &  0.04 &  0.15 &  0.06 \\
  64 & 1431 &  0.96 &  0.03 & \nodata & \nodata &  0.12 &  0.02 & \nodata & \nodata &   63 &  0.16 &  0.26 &  1.00 &  0.02 &  0.16 &  0.02 \\
  65 & 1528 &  0.89 &  0.02 & \nodata & \nodata &  0.12 &  0.03 & \nodata & \nodata &   41 &  0.10 &  0.31 &  0.95 &  0.03 &  0.24 &  0.04 \\
  66 & 1695 &  0.76 &  0.09 &  1.19 &  0.06 &  0.11 &  0.03 &  0.58 &  0.22 &   14 &  0.20 &  0.25 &  1.01 &  0.06 &  0.23 &  0.04 \\
  67 & 1833 &  0.90 &  0.09 & \nodata & \nodata &  0.14 &  0.03 & \nodata & \nodata &   20 &  0.29 &  0.31 &  1.01 &  0.05 &  0.23 &  0.04 \\
  68 &  437 &  0.83 &  0.09 & \nodata & \nodata &  0.12 &  0.03 & \nodata & \nodata &   38 &  0.10 &  0.31 &  0.90 &  0.04 &  0.23 &  0.05 \\
  69 & 2019 &  0.76 &  0.14 &  1.00 &  0.08 &  0.09 &  0.04 &  0.48 &  0.29 &   27 &  0.33 &  0.37 &  0.90 &  0.03 &  0.14 &  0.02 \\
  70 &   33 &  0.79 &  0.12 & \nodata & \nodata &  0.11 &  0.04 & \nodata & \nodata &    5 &  0.78 &  0.36 &  1.01 &  0.16 &  0.28 &  0.10 \\
\enddata
\end{deluxetable*}
}

{\centering
\begin{deluxetable*}{crrrrrrrrrrrrrrrr}
\tablenum{2}
\tablewidth{0pt}
\tabletypesize{\tiny}
\tablecaption{Color Distributions: Bimodal and Unimodal Parameters}
\tablehead{
\colhead{No.} &
\colhead{VCC} &
\colhead{$\mu_{blue}$} & \colhead{$\mu_{blue,err}$} &
\colhead{$\mu_{red}$} & \colhead{$\mu_{red,err}$} &
\colhead{$\sigma$} & \colhead{$\sigma_{err}$} &
\colhead{$f_{red}$} & \colhead{$f_{red,err}$} &
\colhead{$N_{GC}$} &
\colhead{$p$} & \colhead{$p_{err}$} &
\colhead{$\mu_1$} & \colhead{$\mu_{1,err}$} &
\colhead{$\sigma_1$} & \colhead{$\sigma_{1,err}$} \\
\colhead{(1)} &
\colhead{(2)} &
\colhead{(3)} &
\colhead{(4)} &
\colhead{(5)} &
\colhead{(6)} &
\colhead{(7)} &
\colhead{(8)} &
\colhead{(9)} &
\colhead{(10)} &
\colhead{(11)} &
\colhead{(12)} &
\colhead{(13)} &
\colhead{(14)} &
\colhead{(15)} &
\colhead{(16)} &
\colhead{(17)}
}
\startdata
  71 &  200 &  0.69 &  0.11 &  0.89 &  0.08 &  0.07 &  0.03 &  0.63 &  0.28 &   18 &  0.33 &  0.30 &  0.82 &  0.03 &  0.11 &  0.02 \\
  72 &  571 &  0.69 &  0.08 & \nodata & \nodata &  0.09 &  0.02 & \nodata & \nodata &   11 &  0.26 &  0.26 &  0.92 &  0.06 &  0.20 &  0.04 \\
  73 &   21 &  0.75 &  0.10 & \nodata & \nodata &  0.09 &  0.03 & \nodata & \nodata &   17 &  0.20 &  0.31 &  0.88 &  0.05 &  0.20 &  0.07 \\
  74 & 1488 &  0.72 &  0.10 & \nodata & \nodata &  0.10 &  0.05 & \nodata & \nodata &   10 &  0.22 &  0.27 &  0.87 &  0.07 &  0.23 &  0.08 \\
  75 & 1779 &  0.66 &  0.08 & \nodata & \nodata &  0.11 &  0.06 & \nodata & \nodata &    2 &  0.99 &  0.07 &  0.88 &  0.34 &  0.23 &  0.27 \\
  76 & 1895 &  0.69 &  0.19 & \nodata & \nodata &  0.11 &  0.05 & \nodata & \nodata &    6 &  0.56 &  0.39 &  0.95 &  0.11 &  0.26 &  0.08 \\
  77 & 1499 &  0.82 &  0.08 &  1.18 &  0.18 &  0.14 &  0.03 &  0.39 &  0.28 &   27 &  0.41 &  0.32 &  0.93 &  0.04 &  0.21 &  0.03 \\
  78 & 1545 &  0.90 &  0.02 & \nodata & \nodata &  0.17 &  0.03 & \nodata & \nodata &   53 &  0.14 &  0.33 &  0.93 &  0.02 &  0.22 &  0.03 \\
  79 & 1192 &  0.94 &  0.01 &  1.45 &  0.03 &  0.13 &  0.01 &  0.32 &  0.05 &  200 & $< 0.01$ & $< 0.01$ &  1.10 &  0.02 &  0.28 &  0.01 \\
  80 & 1857 &  0.99 &  0.05 & \nodata & \nodata &  0.07 &  0.02 & \nodata & \nodata &    8 &  0.17 &  0.33 &  1.14 &  0.09 &  0.24 &  0.07 \\
  81 & 1075 &  0.82 &  0.16 & \nodata & \nodata &  0.11 &  0.05 & \nodata & \nodata &   20 &  0.14 &  0.31 &  0.93 &  0.04 &  0.19 &  0.05 \\
  82 & 1948 &  0.72 &  0.06 & \nodata & \nodata &  0.05 &  0.02 & \nodata & \nodata &    6 &  0.67 &  0.37 &  0.83 &  0.06 &  0.12 &  0.04 \\
  83 & 1627 & \nodata & \nodata & \nodata & \nodata & \nodata & \nodata & \nodata & \nodata &    0 & \nodata & \nodata & \nodata & \nodata & \nodata & \nodata  \\
  84 & 1440 &  0.87 &  0.04 & \nodata & \nodata &  0.11 &  0.02 & \nodata & \nodata &   31 & $< 0.01$ &  0.02 &  0.98 &  0.04 &  0.25 &  0.04 \\
  85 &  230 &  0.84 &  0.11 & \nodata & \nodata &  0.11 &  0.03 & \nodata & \nodata &   31 &  0.09 &  0.27 &  0.92 &  0.04 &  0.21 &  0.05 \\
  86 & 2050 &  0.79 &  0.06 & \nodata & \nodata &  0.06 &  0.02 & \nodata & \nodata &   13 &  0.22 &  0.28 &  0.89 &  0.06 &  0.17 &  0.10 \\
  87 & 1993 & \nodata & \nodata & \nodata & \nodata & \nodata & \nodata & \nodata & \nodata &    2 & \nodata & \nodata & \nodata & \nodata & \nodata & \nodata  \\
  88 &  751 &  0.80 &  0.01 & \nodata & \nodata &  0.03 &  0.01 & \nodata & \nodata &   15 &  0.05 &  0.20 &  0.85 &  0.04 &  0.12 &  0.04 \\
  89 & 1828 &  0.73 &  0.16 &  0.97 &  0.12 &  0.08 &  0.03 &  0.61 &  0.39 &   20 &  0.30 &  0.41 &  0.88 &  0.03 &  0.11 &  0.03 \\
  90 &  538 &  0.75 &  0.08 & \nodata & \nodata &  0.07 &  0.03 & \nodata & \nodata &    5 &  0.69 &  0.41 &  0.89 &  0.09 &  0.16 &  0.09 \\
  91 & 1407 &  0.96 &  0.06 & \nodata & \nodata &  0.13 &  0.03 & \nodata & \nodata &   49 &  0.12 &  0.27 &  1.02 &  0.02 &  0.18 &  0.03 \\
  92 & 1886 &  0.70 &  0.05 & \nodata & \nodata &  0.04 &  0.02 & \nodata & \nodata &    6 &  0.58 &  0.43 &  0.80 &  0.05 &  0.11 &  0.03 \\
  93 & 1199 &  0.94 &  0.02 &  1.40 &  0.02 &  0.14 &  0.01 &  0.41 &  0.04 &  216 & $< 0.01$ & $< 0.01$ &  1.13 &  0.02 &  0.27 &  0.01 \\
  94 & 1743 &  0.75 &  0.19 & \nodata & \nodata &  0.13 &  0.07 & \nodata & \nodata &    7 &  0.46 &  0.40 &  1.01 &  0.17 &  0.36 &  0.10 \\
  95 & 1539 &  0.90 &  0.07 & \nodata & \nodata &  0.06 &  0.02 & \nodata & \nodata &   34 &  0.13 &  0.21 &  0.97 &  0.02 &  0.12 &  0.04 \\
  96 & 1185 &  0.85 &  0.03 & \nodata & \nodata &  0.12 &  0.02 & \nodata & \nodata &   26 &  0.10 &  0.22 &  0.92 &  0.04 &  0.21 &  0.03 \\
  97 & 1826 &  0.66 &  0.07 & \nodata & \nodata &  0.08 &  0.03 & \nodata & \nodata &    8 &  0.35 &  0.34 &  0.76 &  0.07 &  0.18 &  0.06 \\
  98 & 1512 &  0.63 &  0.11 & \nodata & \nodata &  0.06 &  0.02 & \nodata & \nodata &    9 &  0.27 &  0.33 &  0.82 &  0.06 &  0.15 &  0.04 \\
  99 & 1489 &  0.83 &  0.04 &  1.22 &  0.05 &  0.09 &  0.02 &  0.40 &  0.16 &   16 &  0.12 &  0.20 &  0.98 &  0.05 &  0.22 &  0.03 \\
 100 & 1661 &  0.75 &  0.17 & \nodata & \nodata &  0.09 &  0.05 & \nodata & \nodata &   12 &  0.19 &  0.29 &  0.95 &  0.06 &  0.20 &  0.08 \\
\enddata
\tablenotetext{1}{Running number, sorted by increasing $B$ magnidtude}
\tablenotetext{2}{Number in Virgo Cluster Catalog}
\tablenotetext{3,4}{Mean of blue peak and error, if bimodal}
\tablenotetext{5,6}{Mean of red peak and error, if bimodal}
\tablenotetext{7,8}{Gaussian $\sigma$ for both peaks, and error}
\tablenotetext{9,10}{Fraction of GCs that are red, and error, if bimodal}
\tablenotetext{11}{Total number of GCs and error}
\tablenotetext{12,13}{p-value and error}
\tablenotetext{14,15}{Mean of distribution and error}
\tablenotetext{16,17}{Standard deviation of distribution and error}
\end{deluxetable*}
}

\begin{figure*}
\plotone{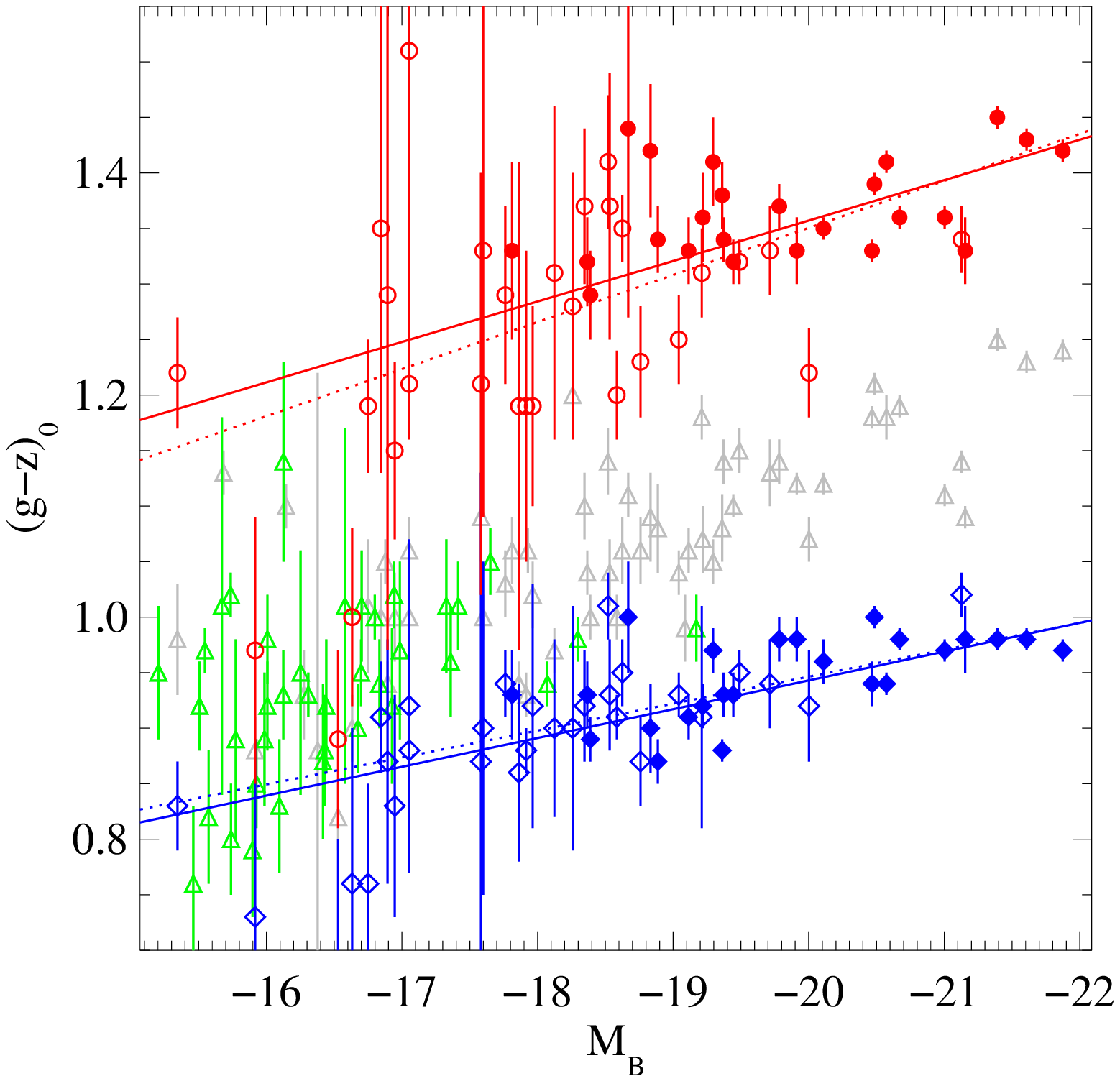}
\caption{Mean colors of the entire GC system and the 
  two GC subpopulations as a function of $M_B$.
  We plot the values of the blue and
  red means as determined by the KMM algorithm assuming a model of two
  Gaussians with a common dispersion.  
  Only 54 galaxies for which the red peak was more than 2-sigma above the
  background are plotted with two peaks (circles and diamonds).  
  Filled symbols represent the 24 galaxies that have significant
  bimodal distributions ($p<0.05$), and the solid lines are fits to
  these points.  Open symbols are for the 30 galaxies with $0.05<p<1$
  and the dotted lines are fit to data of all p-values.
  The means
  of the entire GC color distribution are also plotted (green and gray
  triangles) with 46 of the points (green) representing the GC systems
  determined to be unimodal in color.
  \label{fig:colmagkmm}}
\end{figure*}

\noindent{\bf VCC 1499}:  While this galaxy appears to be an
elliptical on ground based Digitized Sky Survey images, 
our ACS images reveal it to
contain numerous young blue star clusters at the center.  Also, the
color of the galaxy is one of the bluest in the sample, suggesting
that it may be a dI/dE transition object.  In the same field, only
1.4\arcmin\ away, is a true dE, VCC 1491, hence the clusters measured
in this field are a combination of the systems of these two galaxies.

\noindent{\bf VCC 9}: This galaxy has a very low surface brightness
for its reported luminosity.

\noindent{\bf VCC 1938}:  This S0 galaxy is 1.7\arcmin\ from the
neighboring nucleated dE, VCC 1945.  We detect GCs in both galaxies,
although there are many fewer associated with VCC 1945 as it is
$\sim2.5$~mag fainter.

\noindent{\bf VCC 33, 1779, 1627, 1993}: In the full field data for these
galaxies, the number of GC candidates we detect is less than 3$\sigma$
above the expected background.  VCC 1779 shows spiral
structure and dust, is likely to have a younger age, and there are a
two likely star clusters near the center.  The other three galaxies do
not show any obvious concentrations of GCs at their centers although
there may be one or two GCs in each galaxy.  Only for VCC 1627 do we not
detect any GC candidates above the background.

For the purposes of the analysis that follow, we exclude from our
sample the galaxies
VCC 1327, 1297, 1192, 1199 (near giants), 1499, 1779 (younger ages),
and 1938 (two galaxy blend) because 
either the detected globular cluster systems may not be representative
of the targeted galaxy, or the young ages belie the classification as
an early-type galaxy.  This leaves us with 93 galaxies in our sample.

\subsection{GC colors and Host Galaxy Properties}
\subsubsection{Color Decompositions of Individual GC Systems}

\begin{figure*}
\plotone{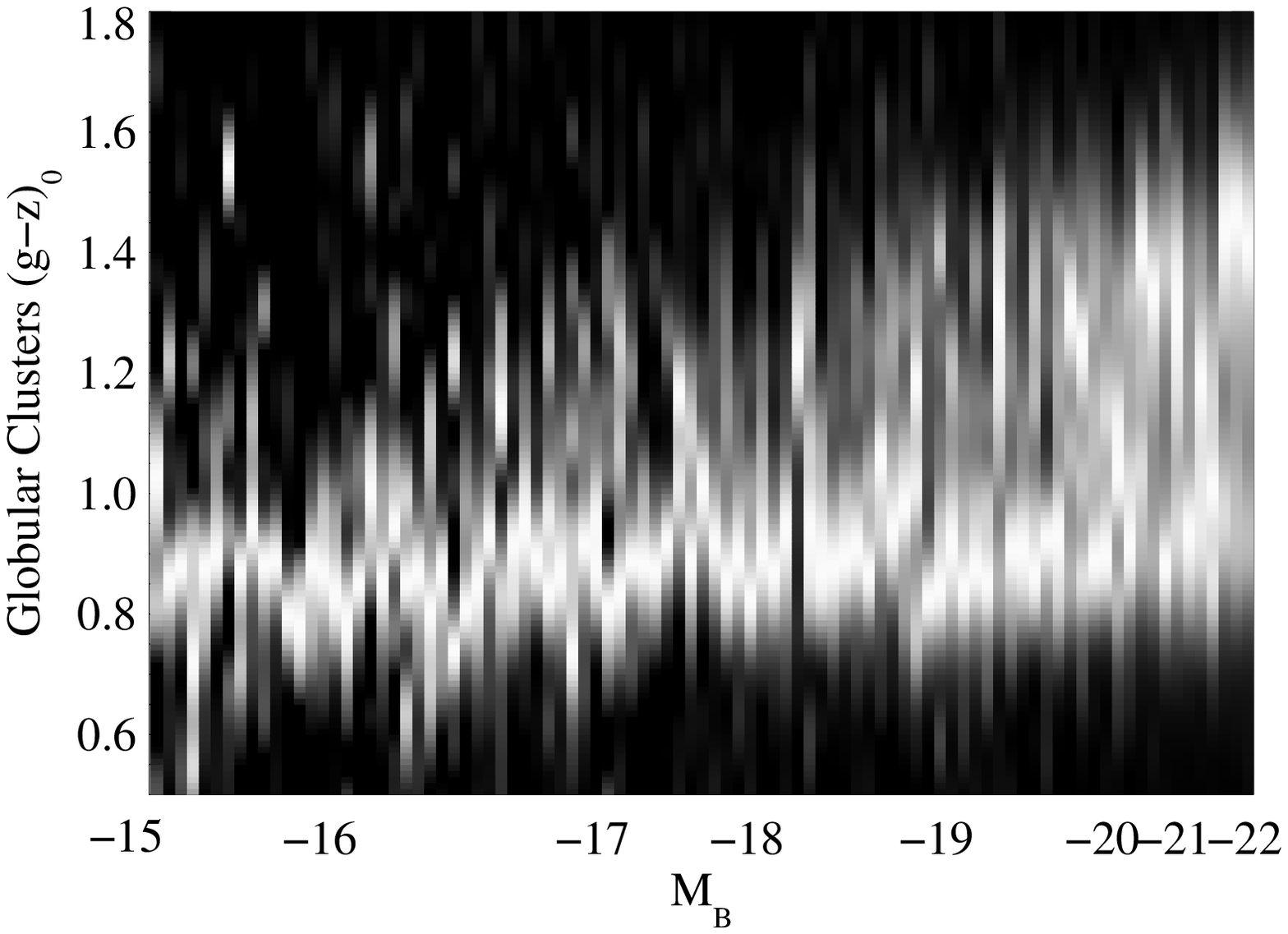}
\caption{This image shows the GC color distributions of the ACSVCS
  galaxies ordered by host galaxy $M_B$.  Each column is a kernel density
  estimation of a galaxy's GC color distribution, with white normalized
  to represent the peak density.  The galaxies are ordered in increasing
  luminosity from left to right.  The image shows that all galaxies seem
  to possess a population of blue GCs with similar color, 
  but that the color and relative
  fractions of red GCs is a strong function of galaxy luminosity.
  \label{fig:colorlum2d}}
\end{figure*}

Using the mixture model estimates presented in
Table~\ref{table:kmmtable}, we can investigate the behavior of the
blue and red GC subpopulations as a function of galaxy luminosity.
Previous studies along these lines were mainly based on HST/WFPC2 data
(Forbes \etal 1997, Kundu \& Whitmore 2001; Larsen \etal 2001;
Burgarella \etal 2001) and
included very few galaxies with $M_B$ fainter than $-19.5$.  
The study of Lotz \etal
(2004) targeted dEs in Virgo and Fornax, and saw no evidence for
bimodality in their GC color distributions.  The higher metallicity
sensitivity of the \gz\ color and the deeper photometry of our ACS
observations makes this an ideal program for testing these relationships
in a homogeneous fashion across a wide range of galaxy luminosities and
colors. 

In Figure~\ref{fig:colmagkmm}, we show the mean colors of the blue
and red GC subpopulations as a function of the absolute blue magnitude
of their host galaxy.  The galaxy magnitudes are from the photometry of
Binggeli, Sandage \& Tammann (1985) 
and are listed for the entire sample in Paper I.  We assume
a distance to the Virgo cluster of 16.5~Mpc (Tonry \etal 2001)
with a distance modulus of $31.09\pm0.03$~mag from Tonry \etal (2001),
corrected by the final 
results of the Key Project distances (Freedman \etal 2001; see also
discussion in Mei \etal 2005b).
The GC color data follow Table~\ref{table:kmmtable}, and
the two individual means are plotted for all galaxies that had a
significant ($>2\sigma$) red GC population.  For 24 galaxies plotted
as solid points, the distributions
were significantly bimodal ($p\leq 0.05$).  
The color distributions for 30 other galaxies (open points) 
can also be decomposed
into two components, but are less uniquely described by a two-Gaussian
model ($0.05 < p \leq 1$).  In both cases, we plot the means of the two
fitted Gaussians only if the number of red GCs is more then
2$\sigma$ above the expected background.  This eliminates small
numbers of red background galaxies from causing spurious bimodality.
For the remaining 46 galaxies (those that have insignificant numbers of
red GCs), we treat them as unimodal and plot the means of their 
entire GC color distributions (triangles).

We are able to resolve the GC
subpopulations for galaxies $\sim2$~mag fainter than those in previous
studies, and we observe a clear
trend that the mean colors of both blue and red GCs are redder for more
luminous host galaxies, and that the slope of the relation for red GCs
is steeper by a factor of 1.6--1.9.  We fit these relations to both
the $p<0.05$ sample and the full sample, deriving the following
weighted linear fits of the form
\begin{equation}
\langle g-z \rangle = a + b\times M_B
\end{equation}
Coefficients for these fits are listed in Table~\ref{table:clfittable}.  
While most galaxies with $M_B<-18$ have distinguishable blue and red
components in their
GC color distributions, the fraction of galaxies whose distributions are
not well resolved increases for fainter galaxies until all of
the galaxies with $M_B > -16$ have an insignificant number of red GCs.  
Because both the fraction of red GCs and the total number
of GCs is lower for fainter galaxies, the red subpopulation becomes more
difficult to separate out at intermediate luminosities.  However, the
slopes that we derive from the brighter galaxies are statistically identical to
those fit to the larger sample.  In other words, even when a galaxy's GC 
color distribution is not statistically very different from unimodal, a
decomposition into two components is still consistent with
the trends seen in more luminous galaxies, and thus may still be an
appropriate description of the system.

\subsubsection{Color Distributions Binned by Galaxy Properties}

\begin{figure*}
\epsscale{1.15}
\plottwo{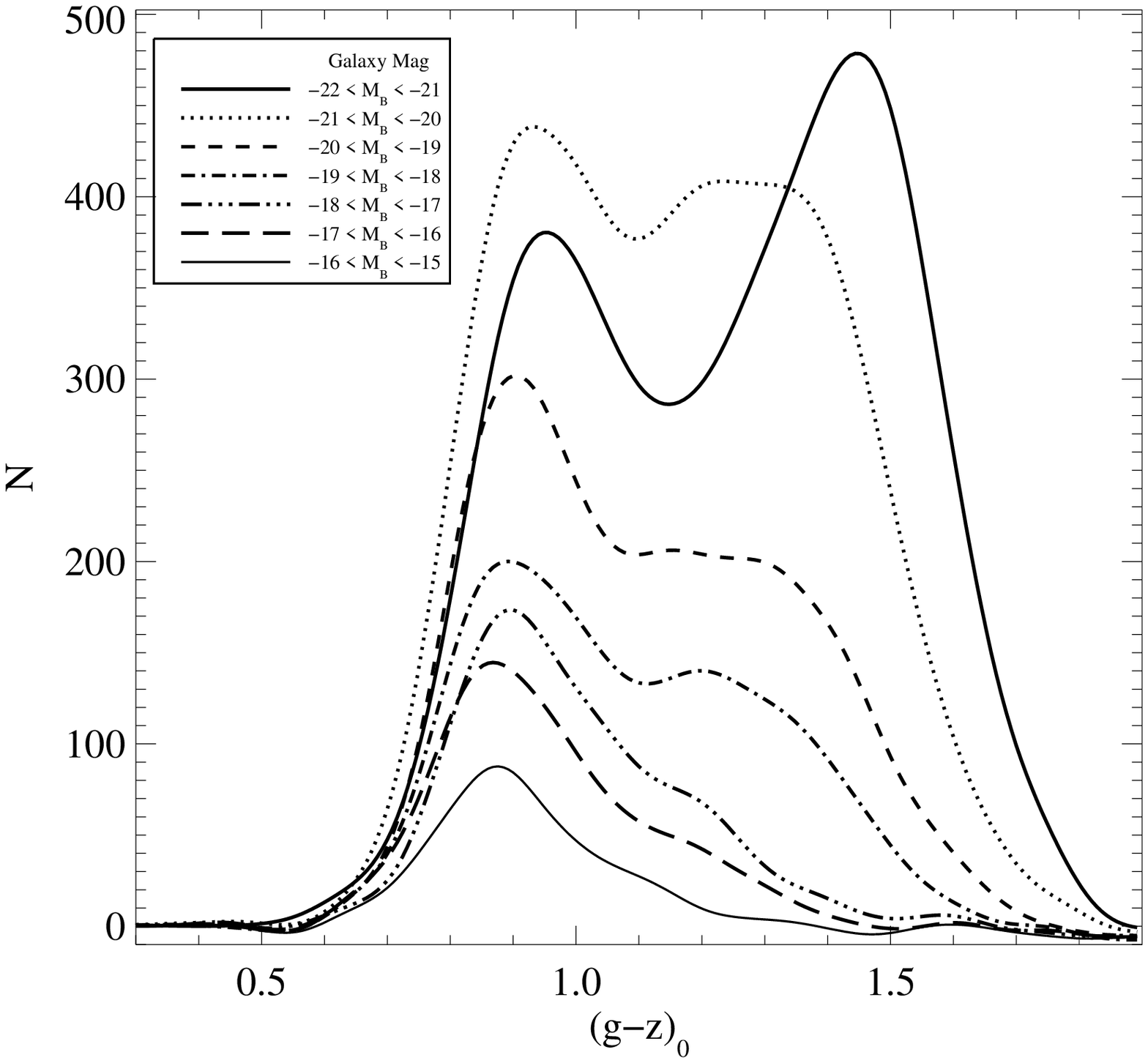}{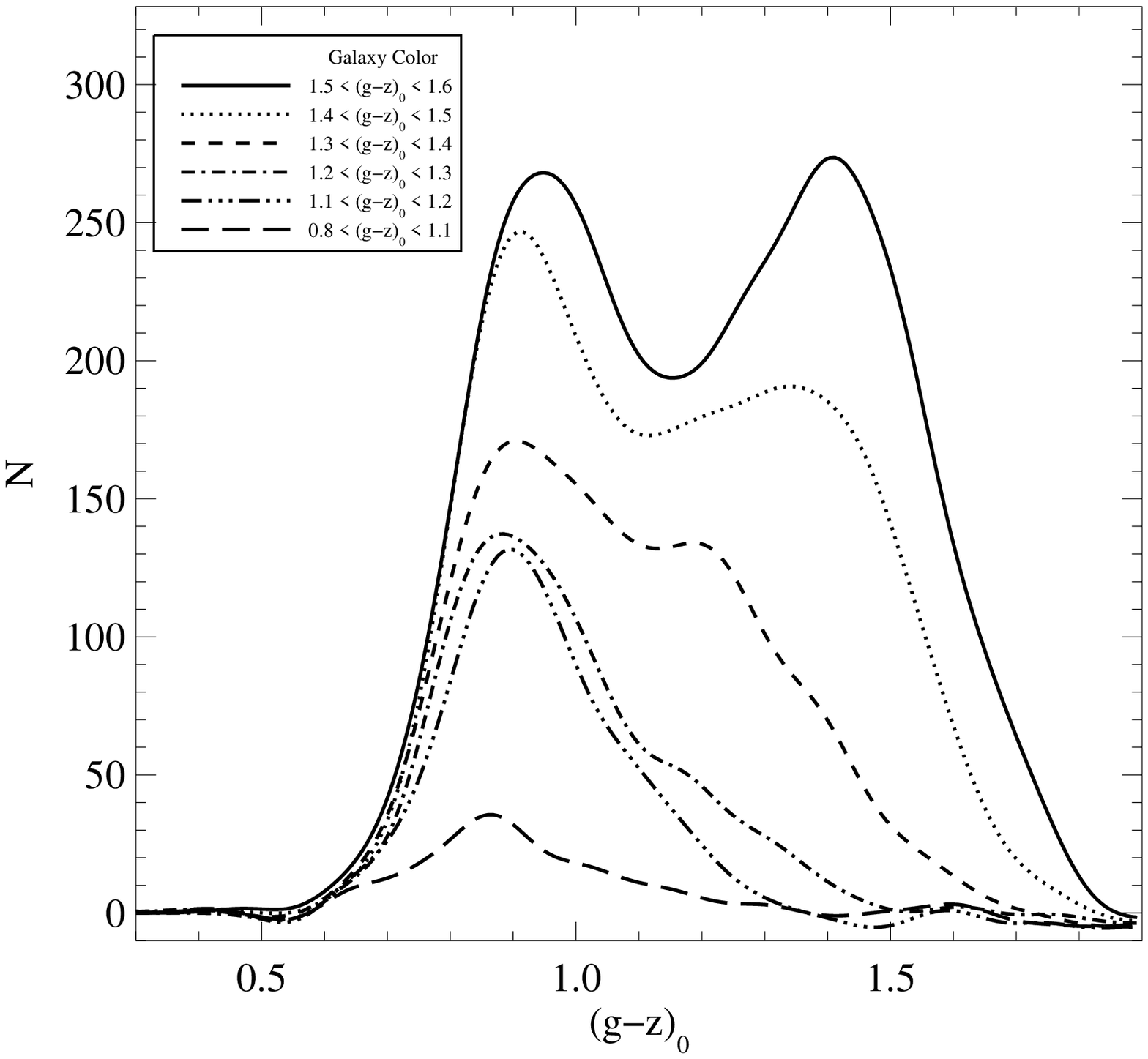}
\caption{Kernel density color distributions of GCs in seven bins of
  host galaxy magnitude (a) and six bins of host galaxy 
  $(g$--$z)_0$ (b).  
  Magnitude bins are 1~mag wide and  extend from 
  $M_B=-21$ to $M_B=-15$ (red to blue).  The distance to the Virgo cluster is
  assumed to be 16.5~Mpc.  Bins of color are 0.1~mag wide from 1.6 to 1.1,
  with the last bin extending to 0.75 (red to blue).  Distributions have been
  corrected for the expected background and foreground contamination as
  measured from our control fields.  All color distributions have been
  constructed with a Gaussian kernel ($\sigma=0.05$~mag).
  Notice how even the faintest and
  bluest bins in our sample appear asymmetric, indicating the presence
  of some red GCs.  Data for this plot is listed in
  Tables~\ref{table:bindecomp_mag}~and~\ref{table:bindecomp_col}
  \label{fig:binkernel}} 
\end{figure*}

When the data is of high enough quality, it is sometimes best to view
the data itself in aggregate rather than struggle with parametrization.
Figure~\ref{fig:colorlum2d} displays a nonparametric representation of
the GC color distributions for 92 galaxies (the 93 from our
analysis sample minus VCC~1627, which has no detected GCs).  Each column of
this plot is the background cleaned color distribution
of a single galaxy's GCs constructed with a Gaussian kernel, 
where the grayscale has been scaled such that
zero density is black and the mode of the distribution is white.  The
galaxies are rank ordered by their luminosity with approximate $M_B$
labeled on the x-axis.  Immediately apparent
from this image are the two GC subpopulations and their behavior with
galaxy luminosity.  Nearly all galaxies appear to possess a blue subpopulation
of GCs, and the mean of this population varies only slowly with galaxy
luminosity.  In addition, there is a population of red GCs whose color
and number fraction increase in a continuous fashion across our entire
sample, spanning seven magnitudes of galaxy luminosity.  Lotz \etal
(2004) made a similar inference when comparing the GCs in dEs to the blue
GCs in ellipticals.  Interestingly, a red wing in
the GC color distribution appears to exist even in some of our faintest
galaxies, although the small number of GCs and 
increased noise due to background subtraction
makes it difficult to quantify this for any individual galaxy.

Figure~\ref{fig:colorlum2d} displays the utility of grouping galaxies
together by their intrinsic properties.
Because of the large number of galaxies in our sample, we can quantify
trends by treating the galaxies collectively even when any single
faint galaxy has too few GCs for substantive analysis.  In
Figure~\ref{fig:binkernel}, we investigate
the GC color distributions as a function of both galaxy luminosity and
galaxy color by binning them and accumulating enough GCs to overcome the
noise.  We create seven bins of magnitude from $M_B = -22$ to $-15$ in steps of
one magnitude.  We have six bins in galaxy color, five from $(g-z)_0 = 1.6$ to
$1.1$ in steps of 0.1~mag, and the last bin somewhat larger $0.75 <
(g-z)_0 < 1.1$.  The colors of the galaxies were derived from our ACS
images as described in Paper VI. Briefly, the colors were
estimated by directly integrating the observed surface brightness
profiles between 1$\arcsec$ and the smaller of one effective radius or
the radius at which the surface brightness profile falls one magnitude
below the sky (in either filter). 

The cumulative background cleaned GC color distributions for these galaxy bins
are presented in Figure~\ref{fig:binkernel}.  The trends seen in
Figure~\ref{fig:colorlum2d} are also evident in the first plot of
Figure~\ref{fig:binkernel}.  In each color distribution
a strong blue peak is easily visible, and each distribution is
asymmetric about this blue peak, even the one for our faintest bin.  The
behavior of the red peak is also easy to discern.  The number fraction
and color of the red peak progressively decreases for fainter galaxies.
Where the red peak is obvious, it also appears to be broader than the
blue peak. The same progression is seen for bins in galaxy color.
Redder galaxies have more red GCs, and these red GCs are themselves redder.

Each of the magnitude and color bins across our entire sample
possess a population of blue GCs with some additional red GCs.  
We can try to decompose these two populations in our binned sample in
a way similar to how we treated the individual galaxies, by treating
them as the sum of two Gaussian distributions.  However, given the
higher signal-to-noise of these summed distribution, we find that the
sum of two Gaussians, whether with identical or independent variances,
is often not a good model for the color distributions for the
purpose of measuring the colors of the two modes.
When we run the KMM
algorithm on these data to test for bimodality, all the color distributions
return small $p$-values, including the faintest and bluest bins.
However, when estimating the mean values of the blue and red GC peaks,
the two-Gaussian fits systematically skew the colors of the blue and
red peaks toward the median of the entire distribution---i.e.\ the
estimated color of the blue peak is too red, and the estimated color
of the red peak is too blue as compared to a non-parametric estimation
from the data.  This effect is most pronounced when the modes are of
nearly equal strength, and the result is to introduce a bias that is
correlated with galaxy luminosity or color.  This is something that
needs to be treated with care because it can produce correlations that
are merely artifacts of the parameter estimation.

\subsubsection{Nonparametric Decompositions of Binned GC Color Distributions}

While the addition of another Gaussian component (and its attending 
free parameters) does improve the fit with marginally significant
$p$-values in some cases, we instead choose to avoid this somewhat
arbitrary model and decompose the two peaks in a
nonparametric way.  Given the ubiquity of the blue GCs, we start
with the assumption that every color distribution is made up of a
population of blue GCs with a symmetric distribution about the blue
mode.  We then take the GCs blueward of the blue mode, reflect them
about the blue mode, and take that to be the blue GC population.
The remainder of the GCs are then what we consider the red
GCs.  The peak color of the red peak is estimated from the GC color
distribution that has the blue GCs subtracted.  While this method does
make assumptions, particularly about the symmetry of the blue GC color
distribution and the choice of a kernel size, unlike a
multi-Gaussian model it makes no assumptions about the shape of the red
GC color distribution.  The results are also not particularly
sensitive to the size of the kernel as long as it is not so small as to
introduce spurious peaks, and not so large as to be comparable to the
half-width of the blue peak.

\begin{figure}
\epsscale{1.2}
\plotone{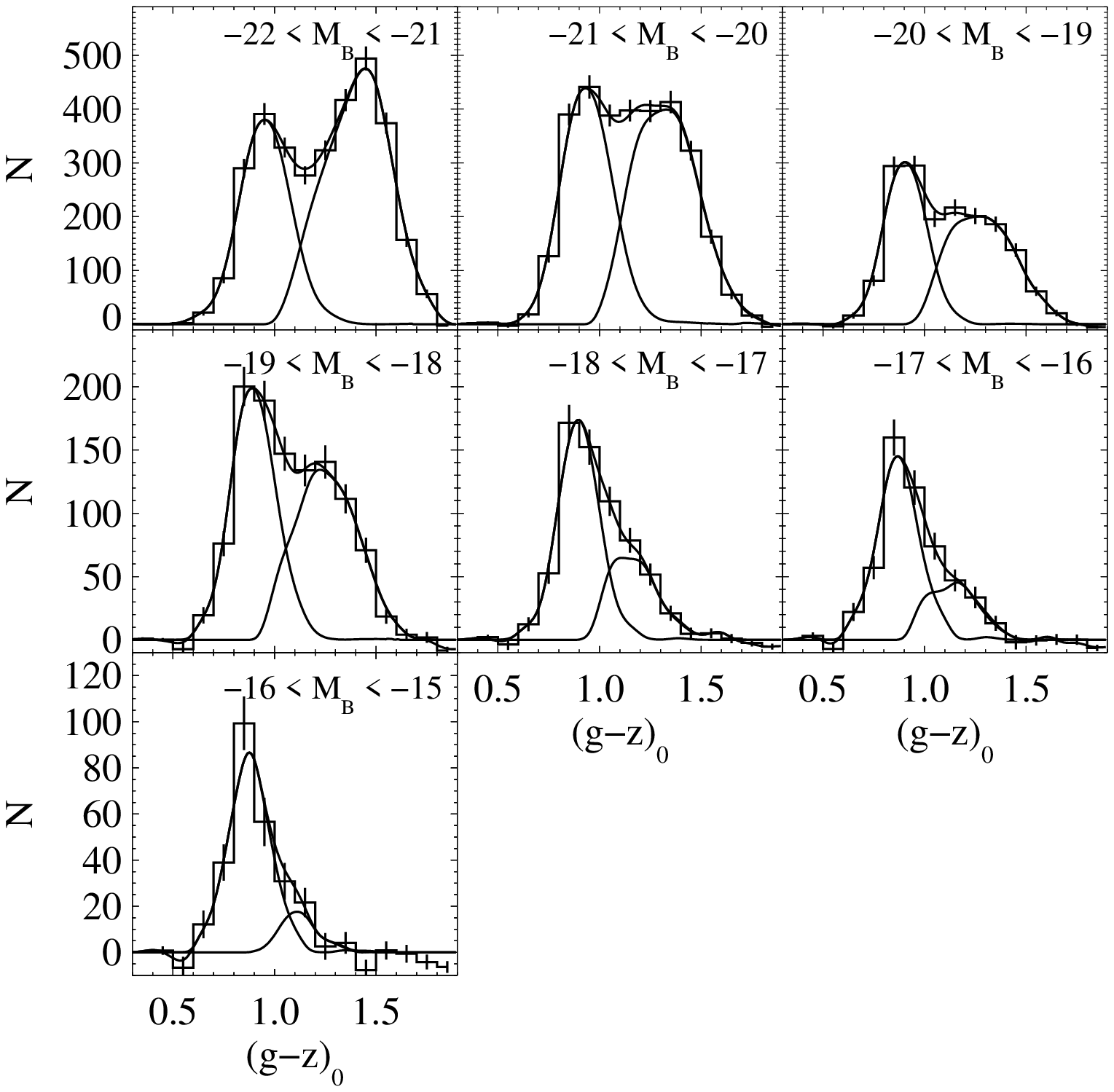}
\caption{Color histograms and nonparametric decompositions of GCs
  binned by host galaxy magnitude.\label{fig:binmag_tg}}
\end{figure}

\begin{figure}
\epsscale{1.2}
\plotone{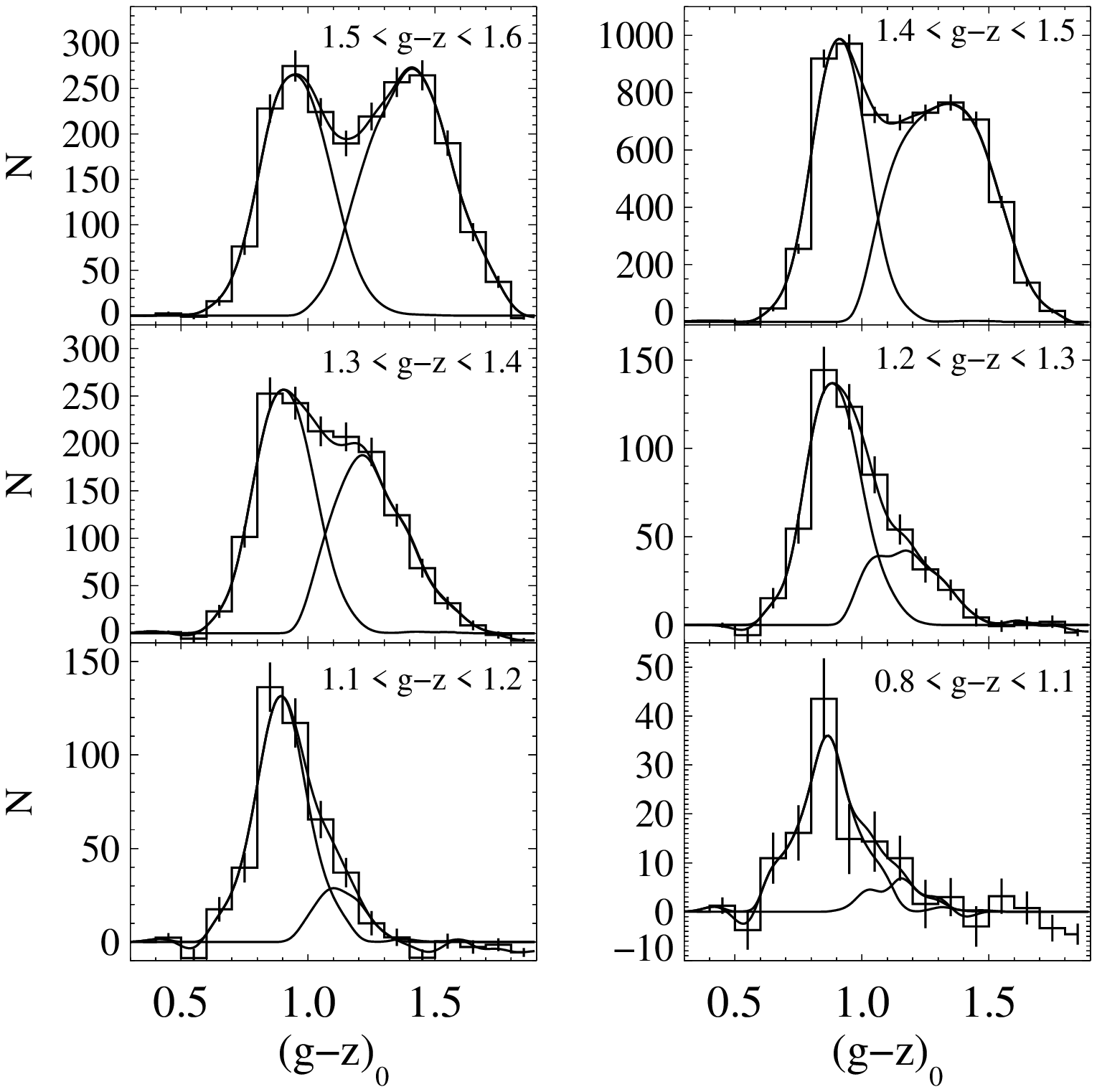}
\caption{Color histograms and nonparametric decompositions of 
  GCs binned by host galaxy color.\label{fig:bincol_tg}}
\end{figure}

In Figures~\ref{fig:binmag_tg} and
\ref{fig:bincol_tg}, we overplot our nonparametric decompositions for
the GC color distributions, binned by galaxy magnitude and color.
We estimate the median
color and the half-width that encompasses 68\% of the GCs for each
peak, and also for the entire GC population in that bin.  The
parameters and their errors for this model were estimated using the
bootstrap with 1000 iterations.  These values are listed in
Tables~\ref{table:bindecomp_mag} and \ref{table:bindecomp_col}.

\begin{figure}
\epsscale{1.22}
\plotone{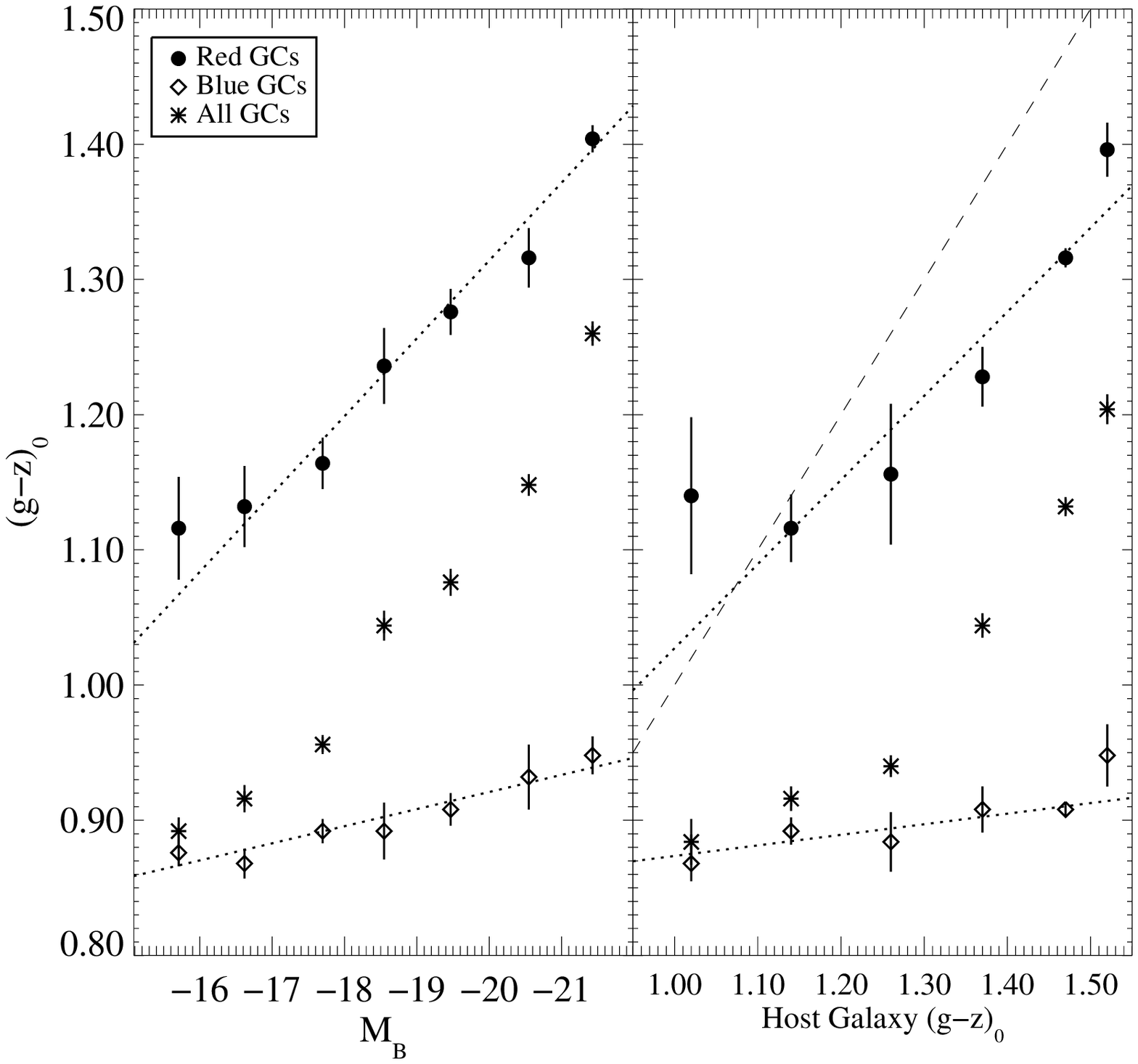}
\caption{Mean values of blue, red, and total GC colors as a function of host
  galaxy luminosity (left) and color (right).  Magnitude and color bins,
  are the
  same as in Figure~\ref{fig:binkernel}.  
  Points represent the blue peak of
  the GC color distribution (blue diamonds), the red component (red
  filled circles), and the mean of the entire distribution (asterisks).
  The dotted lines are linear fits to the two subpopulations, and the
  dashed line (right) has slope unity and represents the galaxy colors.
  Both populations of GCs get
  redder with galaxy luminosity and color, but the slope for the red GCs
  is many times steeper. \label{fig:bincolmag}}
\end{figure}

\begin{figure}
\epsscale{1.22}
\plotone{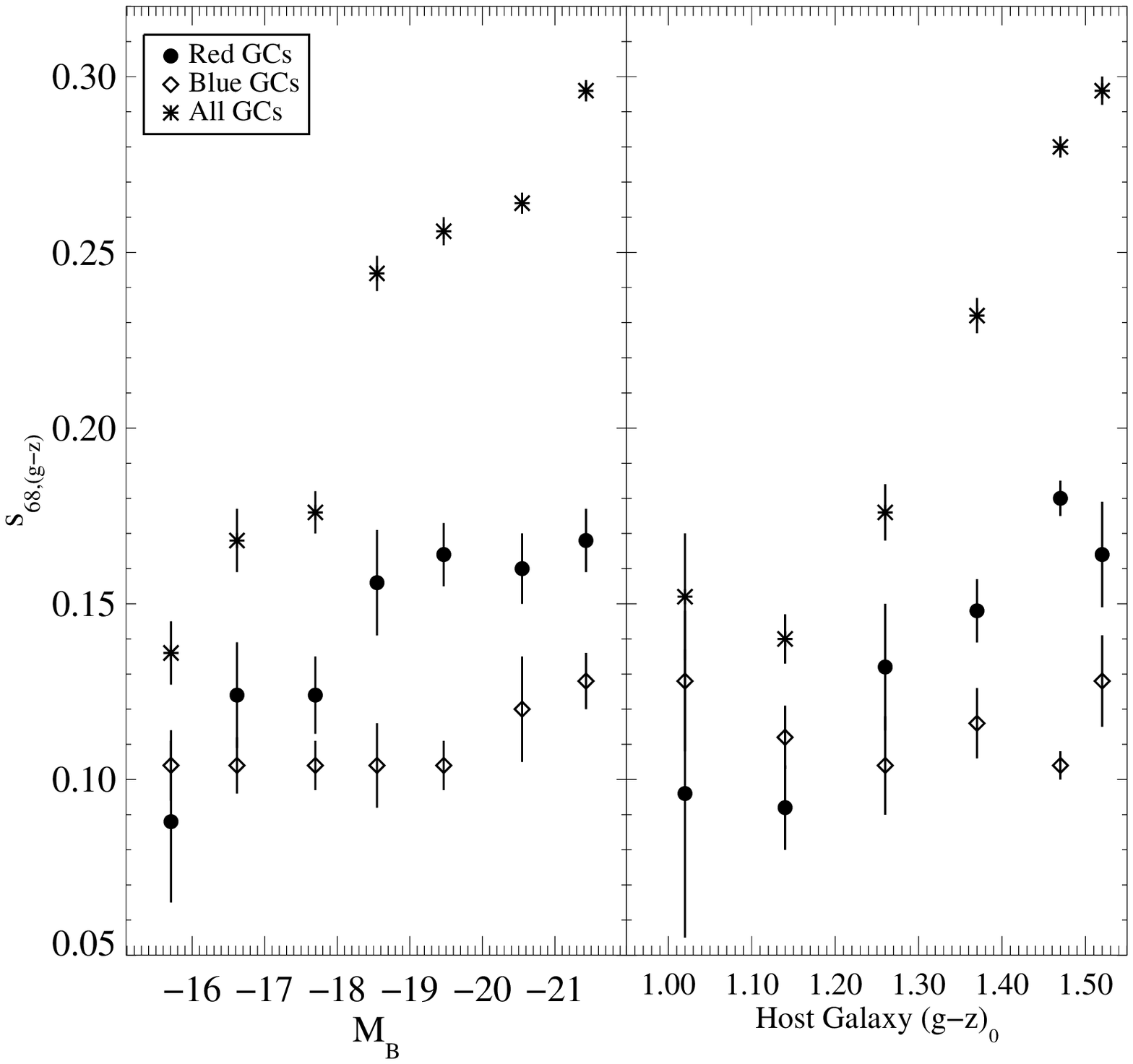}
\caption{Color dispersions ($s_{68}$, the 68\% half-width) 
  of blue (diamond), red (dot), and total (asterisk) GC
  distributions as a function of host galaxy luminosity (left) and color
  (right) binned by magnitude. Magnitude, and color bins
  same as in Figure~\ref{fig:binkernel}.  For luminous and red galaxies,
  the width of the red subpopulation appears to be larger that that
  for the blue GCs.
  \label{fig:binsig}}
\end{figure}

\begin{figure}
\epsscale{1.22}
\plotone{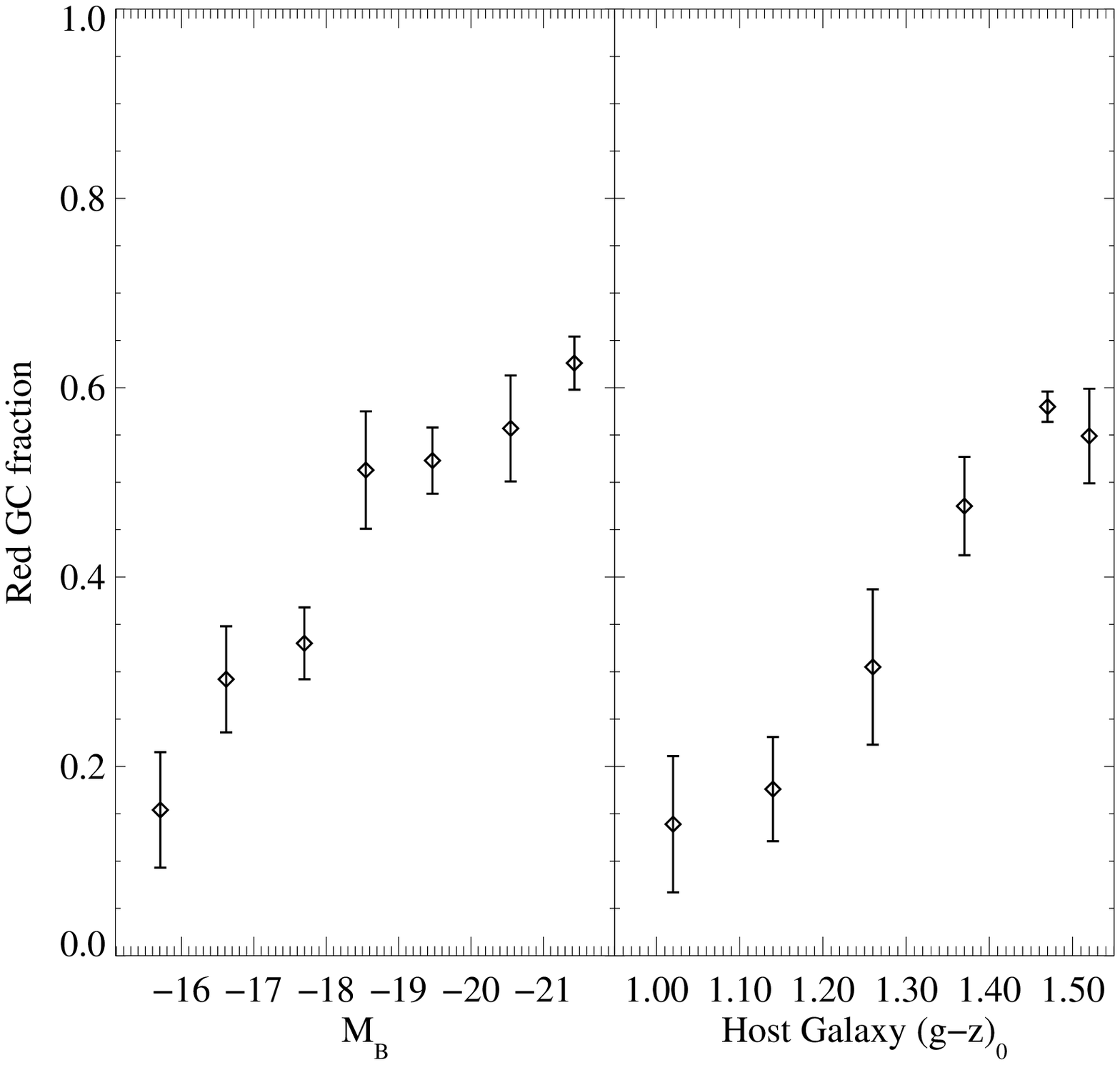}
\caption{Fraction of red GCs as a function of host galaxy luminosity,
  binned by magnitude.  The fraction of red GCs increases by over a
  factor of six across the magnitude range of our sample. 
  \label{fig:redfrac}}
\end{figure}

Using these decompositions, we can study the properties of the GC
color distribution as a function of galaxy properties.
The results of these fits are shown in
Figures~\ref{fig:bincolmag}-\ref{fig:redfrac}.  In
Figure~\ref{fig:bincolmag}a we show the relationship between the median
colors of the red, blue, and total GC populations against galaxy
luminosity and galaxy color.  As in Figure~\ref{fig:colmagkmm}, the colors
of the red, blue, and total GC populations correlate with the
luminosity of the host galaxy.  The coefficients for the linear fits are
presented in Table~\ref{table:binfit}.  The relationship versus luminosity is
similar to that shown with individual galaxies in
Figure~\ref{fig:colmagkmm}.  However, there are two notable
differences.  First, we are able to trace the mean colors of the red and
blue peaks for the entire magnitude range of our sample with much less
noise.  Second, the slopes of the red and blue relations are more
disparate than they are in Table~\ref{table:clfittable}.  
The slope for the red GCs is steeper and the slope for the
blue GCs is shallower than was derived from individual galaxies.  The
result is that the slope for the red GCs is 4.6 times steeper than for
the blue GCs.

Why would the two different methods give different values?
This difference stems from two effects.  
The first is that the individual galaxy
and binned galaxy data were decomposed using two different methods.  When
we apply the KMM algorithm to the binned data, we do get a steeper
blue GC relation and a shallower red GC relation.  However, as we
noted before, this is at least in part due to biases in the fitting of
the peaks that arises from non-Gaussianity in the underlying
distributions.  While a homoscedastic two-Gaussian fit is often the best one
can do for the GCS of a single galaxy, the binned data provide a more
critical test of an inadequate model, and also make nonparametric
methods more feasible.  

The second reason for the difference is that the GC systems of
individual galaxies 
sometimes cannot be decomposed into two populations, either  because
they have no red GCs, the number of red GCs has low
significance, or the total number
of GCs is simply insufficient.  These galaxies are necessarily
dropped, thus biasing the data toward galaxies
with well-spaced GC subpopulations.  This is a bias that will be
strongest at the faint end of the galaxy luminosity function, and will
result in a shallower slope for the red GCs and a steeper slope for
the blue GCs.  The binned data includes all galaxies, regardless of
their individual decomposition, and thus is less biased in this fashion.

Figure~\ref{fig:bincolmag}b also shows the relationship between the
colors of these peaks with the color of the galaxy.  The bluest bin is
somewhat suspect as it contains few galaxies and is quite noisy.  The
other bins, however, each contain $\gtrsim400$~GCs.  Both of the GC
subpopulations show a clear correlation with galaxy color, with the
red GCs showing a steeper relation.  Larsen \etal (2001) found that the
two peaks correlate with galaxy color at the 2--$4 \sigma$ level (see
their note added in manuscript), and the higher precision of the ACS
photometry for both the GCs and the galaxies leaves little to doubt.
While this relationship might be expected since galaxy color is known to
correlate with luminosity, the color-magnitude relationship for galaxies
has significant scatter at the faint end (see Ferrarese \etal 2005) 
and it is not obvious that a tight correlation
should necessarily exist.  The coefficients of the weighted linear
fits are given in Table~\ref{table:binfit}.
The dashed line in this plot shows the one-to-one relation between host
galaxy and GC colors.  For bright galaxies, the median color of the
entire GC system appears to track the galaxy color closely with a
$\sim0.3$~mag offset, but the slope changes for faint galaxies.

In Figure~\ref{fig:binsig}, we show the fitted color dispersion in
each subpopulation, and the dispersion of
the entire color distribution as a function of galaxy luminosity and
galaxy color.  In both plots, we can see that the dispersion in the GC
colors increases for more luminous and redder galaxies.
Although most work on 
individual galaxies necessarily assume that the widths of the blue and red
distributions are the same, these decompositions for our
binned distributions shows that, at least for the brighter galaxies, the
dispersion in color of the two populations may not in fact be the same
but that the red GCs may have a larger spread in color.  We note that
the way we have chosen to do our nonparametric decomposition, the
width of the blue GC distribution is entirely dependent on the half
of that is blueward of the peak.  However, KMM
estimates of the dispersions using a heteroscedastic Gaussian model
give a similar result.

One of the other noticeable trends across galaxy luminosity and color
involves the fraction of red GCs.  Figure~\ref{fig:binkernel} shows how
the fraction of red GCs is much higher in more luminous and redder
galaxies.  We quantify this in Figure~\ref{fig:redfrac}, showing that
even in the faintest bin in our sample, galaxies on average 
have a $\sim15\%$ 
fraction of red GCs, and this increases to $\sim60\%$ for the most luminous
galaxies.  In the most luminous galaxies, however, we are only
sampling the inner regions of the GC system and color gradients could
slightly affect the total fractions (see e.g.\ Rhode \& Zepf (2004) for a
wide-field study).  Nevertheless,  
This changing fraction of red GCs is the main driver behind
the correlation between the mean color for the entire system and the
luminosity of host galaxy.

\section{Discussion}
\subsection{The \gz--[Fe/H] Relation for Globular Clusters}
In order to make inferences on the nature of GCs and their host
galaxies, it is necessary to translate observables (color, magnitude)
into physical properties (metallicity, luminosity, mass).  One of the
more critical transformations is that of color to metallicity for
globular clusters.  Because GCs are old, simple stellar populations,
broadband color should be a good proxy for metallicity.  Beyond an age
of a few Gyr, the color of a star cluster is predominantly determined
by its metal content, and because it was formed in a single burst,
there is no complicated star formation history to disentangle.
However, in practice, determining the precise transformation from
different filter systems to [Fe/H] even for this simplest of stellar
populations is a difficult task.

Evolutionary synthesis models of stellar populations can in theory
produce these relations, but broadband colors provide a challenge as 
they are highly dependent on the spectrophotometry of the input stellar
libraries, whether theoretical or empirical.  Recent models (e.g.\
Bruzual \& Charlot 2003) show reasonable agreement with Milky
Way, M31, and Magellanic Cloud star clusters.  Others have taken an
empirical approach and fit various relationships to the measured
colors and metallicities of GCs in the Milky Way and other nearby
systems.  The color used most often is $V$--$I$ and numerous
relations have been calibrated, although their slopes can vary by a
factor of two (Couture, Harris, \& Allwright 1990; 
Kissler-Patig \etal 1997; Kissler-Patig \etal 1998; Kundu \& Whitmore
1998; Barmby \etal 2000).  All of these calibrations rely heavily on
the Milky Way GCs (although Kissler-Patig 1998 includes GCs from NGC
1399 to supplement the metal-rich end).  The main limitations of
this approach are the homogeneity of the Galactic GC photometry, the
accuracy of the published reddenings, and the inherent sensitivity of
the $V$-$I$ color to metallicity.  Moreover, models predict that the
color-metallicity relationship should be nonlinear, especially for
metal-poor populations where metallicity sensitivity in the optical is
diminished.

\begin{figure}
\epsscale{1.25}
\plotone{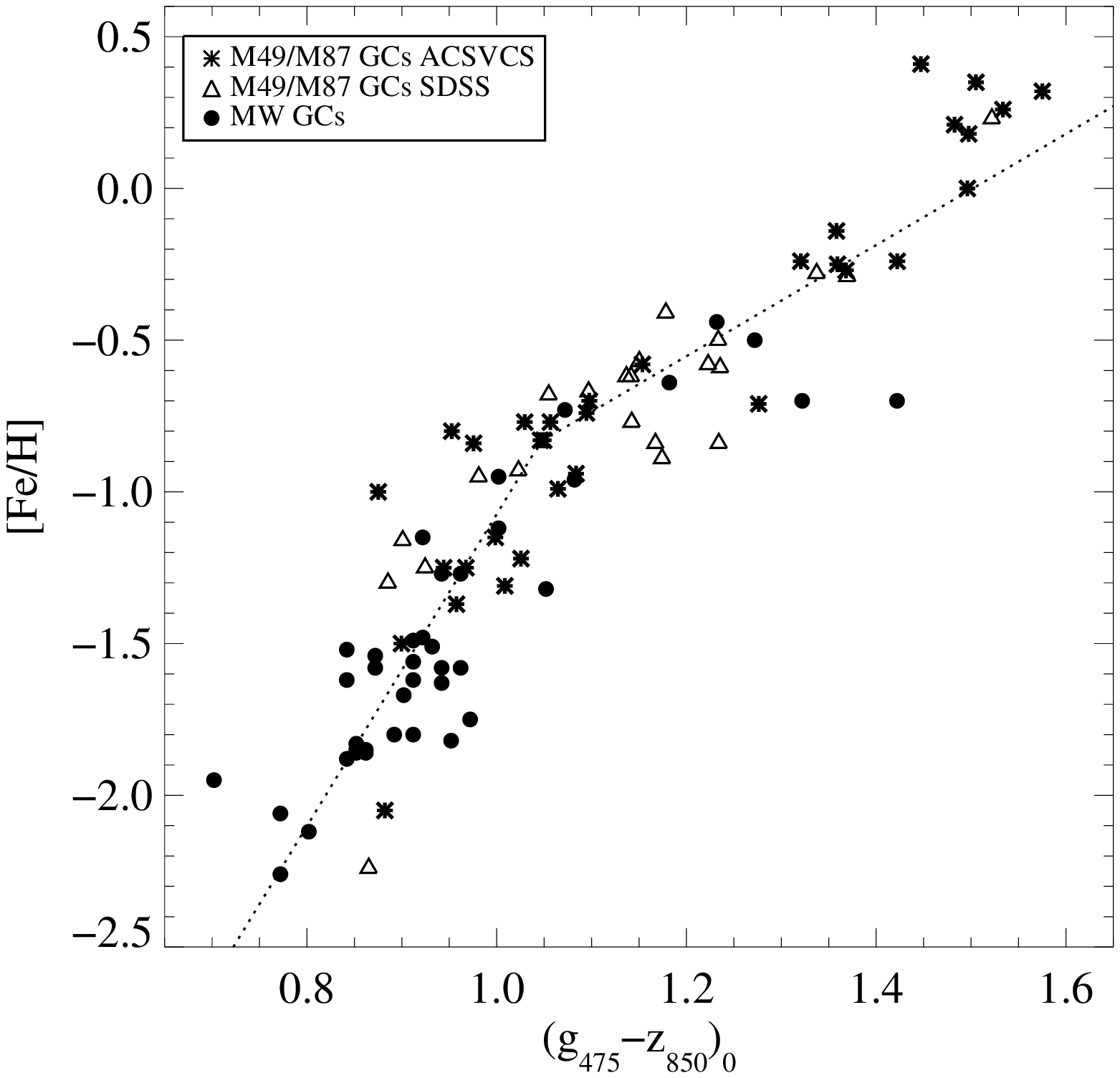}
\caption{[Fe/H] versus $(g$--$z)_0$ for 40 low-extinction
  Milky Way GCs,
  33 M49 and M87 GCs with ACSVCS photometry, and 22 M49 and M87 GCs
  with SDSS photometry.  All GCs have published spectroscopic metallicity
  determinations as described in the text.  The relatively low scatter
  in the data compared to previous work shows that the relation is
  very likely non-linear, with a steepening at ${\rm [Fe/H]} <-0.8$.  The
  dotted line is a broken linear bisector fit to all GCs with
  ${\rm [Fe/H]}<0$.  \label{fig:fehgz}}
\end{figure}

The color that we are working with, \gz, is twice as sensitive to
metallicity as $V$-$I$ (Paper I), and thus offers higher metallicity
resolution than previous studies.  Because of the importance of
deriving this transformation, we have completed a program to image
the Milky Way GC system in these two bandpasses in order to derive an
empirical \gz--[Fe/H] relation.  We used the Cassegrain Focus CCD
Imager on the CTIO 0.9-meter telescope over two observing runs (8 May
-- 12 May 2003 and 31 May -- 6 June 2004) to observe $\sim100$ Milky
Way GCs in the $g'$ and $z'$ filters.  We present a
preliminary color-metallicity 
relation in this section, while the details and final analysis of
this data set will be presented in a separate paper (West et~al., in
prep).

From a sample of 76 Milky Way GCs with good photometry, 
we selected 40 that have low
reddenings of $E(B-V)<0.3$.  Reddenings and [Fe/H] values for these GCs
were obtained from the McMaster Milky Way GC catalog (Harris 1996)
with reddenings and metallicities predominantly compiled from Reed,
Hesser, \& Shawl (1985), Webbink (1985), Armandroff \& Zinn (1988),
and Zinn (1985).  To supplement this sample, especially at
higher metallicities, we added GCs in the giant ellipticals M87 and M49
that have spectroscopic metallicities (Cohen, Blakeslee, \&
Ryzhov 1998; Cohen, Blakeslee, \& \cote 2003).  The latter metallicities
were measured using models of Worthey (1994) and calibrated to the
metallicity scale of Zinn \& West (1984).  For 33 of these GCs, we
have \ggzz\ colors from our ACS/WFC photometry.  For another 22,
we were able to retrieve $g_{SDSS}$ and $z_{SDSS}$ 
photometry from the Sloan Digital
Sky Survey Third Data Release (Abazajian \etal 2005), and correct
them for reddening from the maps of Schlegel \etal (1998).  All photometry
was shifted to the HST/WFC AB photometric system ($g_{475}$ and $z_{850}$).

In Figure~\ref{fig:fehgz}, we show the combined sample of 95 globular
clusters with both a metallicity determination and a \gz\ color.  
While there is still sizable scatter in the data, the relation is
tight enough to show a significant departure from linearity with the
relation steepening for [Fe/H]$<-1.0$.  The dotted line shows a broken
linear fit to the data, with a break point of \gz$=1.05$.  
Because there is appreciable scatter
in both axes, we fit the relation as both [Fe/H] vs.\ \gz\ and \gz\
vs.\ [Fe/H], and take the mean of the results (e.g.\ Barmby \etal 2000).
We take this approach because there is both intrinsic and
observational scatter in both variables, and the scatter is not
well-quantified.  We also fit the data using an ordinary least squares
bisector fit and the method of bivariate
correlated errors with intrinsic scatter (BCES; Akritas \& Bershady
1996).  We find that both of these methods give similar results, with
the slopes varying by less than $4\%$ depending on the chosen
method, a range much smaller than the error (see also
Isobe \etal 1990 for a discussion of linear regression methods).  
Our fit does not include the GCs with [Fe/H]$>0$.  Not only are the
Worthey (1994) models not well-calibrated in that regime, but the
mean colors of the red GCs that we are concerned with in this paper
always have \gz$_0<1.5$. 

The relation we derive is:

\begin{equation}
{\tiny
 {\rm [Fe/H]} = \left\{ \begin{array}{ll}
  -6.21 +  (5.14\pm0.67) (g_{475}-z_{850}), & \mbox{$0.70<(g_{475}-z_{850})\leq1.05$} \\
  -2.75 +  (1.83\pm0.23) (g_{475}-z_{850}), & \mbox{$1.05<(g_{475}-z_{850})<1.45$}
\end{array}
\right.
\label{eqn:gzfeh}
}
\end{equation}

\begin{figure}
\epsscale{1.25}
\plotone{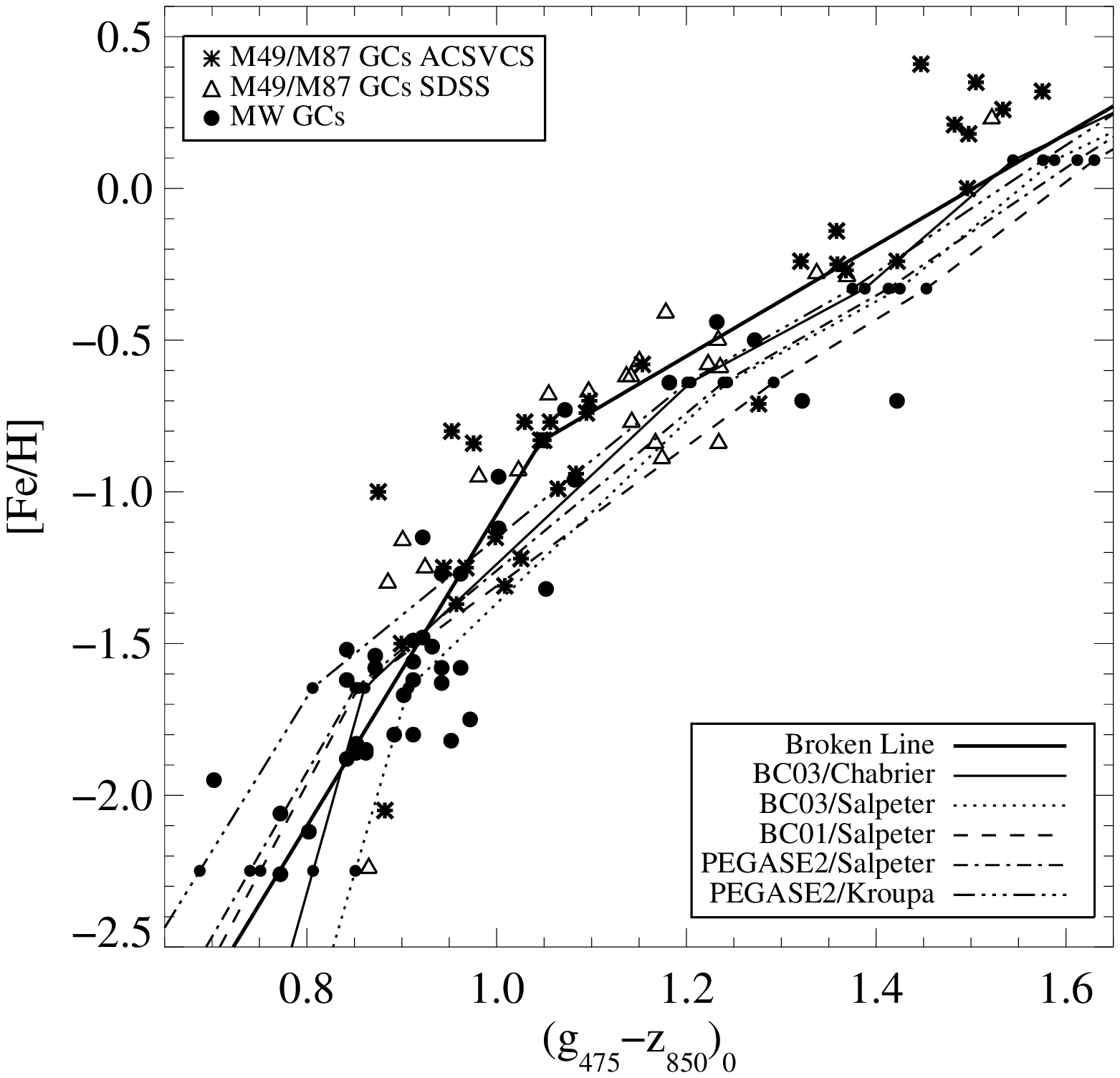}
\caption{[Fe/H] versus $(g$--$z)_0$ for Milky Way, M49, and
  M87 GCs (as in Figure~\ref{fig:fehgz}) with stellar population
  models.  For each model we plot the [Fe/H] and $(g$--$z)_0$ values
  for a 13~Gyr simple stellar population (black points, linearly
  interpolated).  The solid purple line is
  the broken linear bisector fit from Figure~\ref{fig:fehgz}.  While
  the models show decent agreement with the data, there is a large
  spread at low metallicity, and there are also no models for 
  metallicities in the crucial range $-0.6<{\rm [Fe/H]}< -1.6$.
  \label{fig:fehgzmodels}}
\end{figure}

In Figure~\ref{fig:fehgzmodels}, we show the same data and the broken
linear fit, but plot predicted colors from three different models with
different initial mass functions (IMF).  We choose the high resolution
models of 
Bruzual \& Charlot (2003) with IMFs of Chabrier (2003) and Salpeter
(1955), the 2001 release of the Bruzual \& Charlot (1993) models, and
the P{\'E}GASE models, Version 2 (Fioc \& Rocca-Volmerange 1997) with
both Salpeter and Kroupa (2001) IMFs.  In all cases, we have linearly
interpolated the model data (black points) for a 13~Gyr old simple stellar
population.  While there is decent agreement between the models
at metallicities [Fe/H]$>-1$, the relations are increasingly in
disagreement at lower metallicities, and the steepness of the relation
makes the choice of IMF an important one.  Additionally, the lack of
model isochrones for $-1.6 < {\rm [Fe/H]} < -0.7$ means
that there is little information on the color-metallicity relation for
a range that is crucial for GCs, and where the slope of the relation
is rapidly changing.  Because of this, we will adopt our preliminary empirical
relation to obtain metallicities in this paper.  Nevertheless, we urge
caution to all who would transform GC colors into
metallicities using either theoretical or empirical relations.

Lastly, we note that by using the Milky Way GCs to calibrate the
color-metallicity relation, we are assuming that all globular clusters
have old ($>10$~Gyr) ages.  This, however, may not be the case in all GC
systems.  For example, spectroscopic age determinations of metal-rich
GCs in a few other galaxies show that some of them can have intermediate
ages of 3--8~Gyr (Goudfrooij \etal 2001; Peng \etal 2004; 
Puzia \etal 2005).  For the trends we
see in color to be caused solely by age trends, though, would require
that the red GCs in our faintest galaxies have a mean age of 3~Gyr,
assuming that the GCs in the most massive galaxies were 13~Gyr old.
While this mean age is unlikely given the current data, it is not out of
the question that smaller galaxies may have somewhat younger metal-rich
GCs. The age difference necessary for the blue GCs would be smaller
(7~Gyr old in our faintest bin).  However, all spectroscopic evidence 
to date indicates that metal-poor GCs in nearby galaxies 
have ages that are consistent with those of the Galactic GC
system.  Thus, we will assume that the Milky Way GCs can provide a good
calibration for extragalactic GCs with the caveat that if there is a
trend for less
luminous or bluer galaxies host younger GCs, it will serve to flatten the
relations we derive for metallicity.

\subsection{Globular Cluster Metallicities, Galaxy Luminosity, 
  and Stellar Mass}

\begin{figure}
\epsscale{1.2}
\plotone{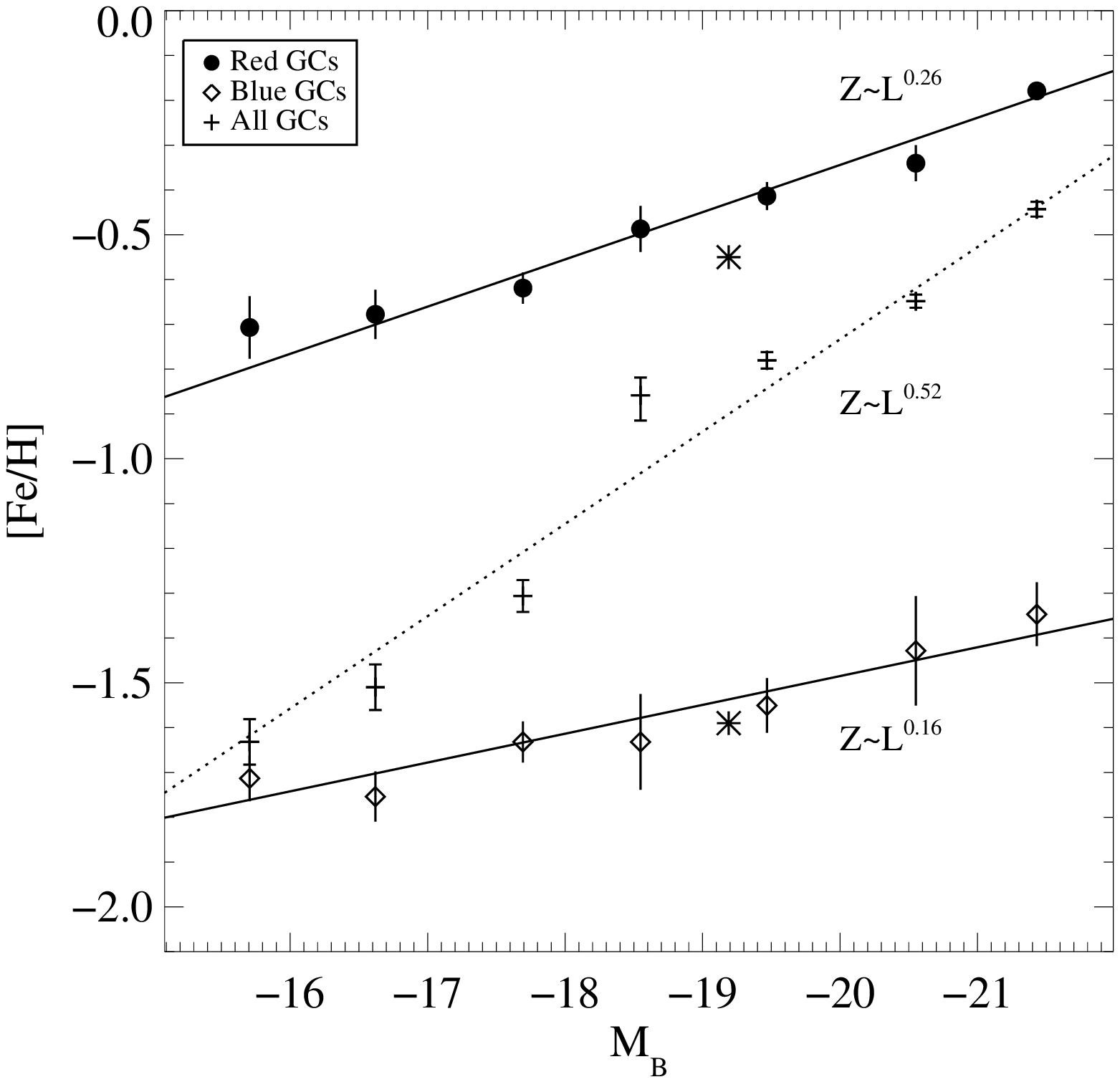}
\caption{Mean metallicities of blue, red, and total GC populations as
  a function of galaxy luminosity.  Bins and points are the same as
  for Figure~\ref{fig:bincolmag}.  The asterisk represents the
  spheroidal component of the Milky Way (bulge and halo) and its GC
  system.  \label{fig:binfehlum}} 
\end{figure}

Using our empirical color-metallicity relation, we can transform the
colors of the binned GC populations to [Fe/H].  In
Figure~\ref{fig:binfehlum}, we plot the metallicities of the blue,
red, and total GC populations against the host galaxy $M_B$.  
The coefficients for the linear fits to these relations are presented in
Table~\ref{table:binfit}.  As was
suggested by Figure~\ref{fig:bincolmag}a, the metallicities of both the
metal-rich and metal-poor GC populations increase for more luminous host
galaxies. One noticeable difference from Figure~\ref{fig:bincolmag}a,
however, is that the slope of the relation for the metal-poor GCs is
more pronounced, and is {\it similar to the slope for the metal-rich GCs}.
Given the \gz-[Fe/H] relation in equation~\ref{eqn:gzfeh}, the
metallicity of the GC populations are proportional to a power of
the luminosity as $Z\propto L^\alpha$ where 
$\alpha = 0.16\pm0.04,\ 0.26\pm0.03,\ {\rm and}\ 0.52\pm0.02$ for the
metal-poor, metal-rich, and total GC populations.  These errors include
the formal errors in the color-metallicity relation.
Lotz \etal (2004) fit the relation to the $V$--$I$ colors of the
total GC populations in Virgo and Fornax dEs (assuming only a single
component), and their slopes were
$Z\propto L^{0.16\pm0.05}\ {\rm to}\ L^{0.22\pm0.05}$ depending on whether they
included Local Group dwarf spheroidals or the blue GCs from Larsen \etal
(2001). This is in reasonable agreement with our fits
especially considering that our sample galaxies are generally more massive, 
and that they used both a different color and color-metallicity relation.

One of the most fundamental galaxy properties is its mass.  While we do
not have dynamical masses for these galaxies yet, we can obtain a rough
estimate of its stellar mass-to-light ratio from its broadband
optical-infrared colors.  We use our \gz\ colors and $J$-$K$ from the
Two Micron All Sky Survey Extended Source Catalog.  These colors and
magnitudes are listed in Paper VI.  Although these
galaxies likely have complex star formation histories that complicate
the determination of $M/L_B$, we can obtain a crude estimate by
comparing their colors to simple stellar population model grids.  We use
the Bruzual \& Charlot (2003) models to obtain an average
luminosity-weighted $M/L_B$ for each galaxy.  We then create seven logarithmic
bins of mass and nonparametrically decompose the GC populations.  These
values are listed in Table~\ref{table:bindecomp_mass}.

\begin{figure}
\epsscale{1.2}
\plotone{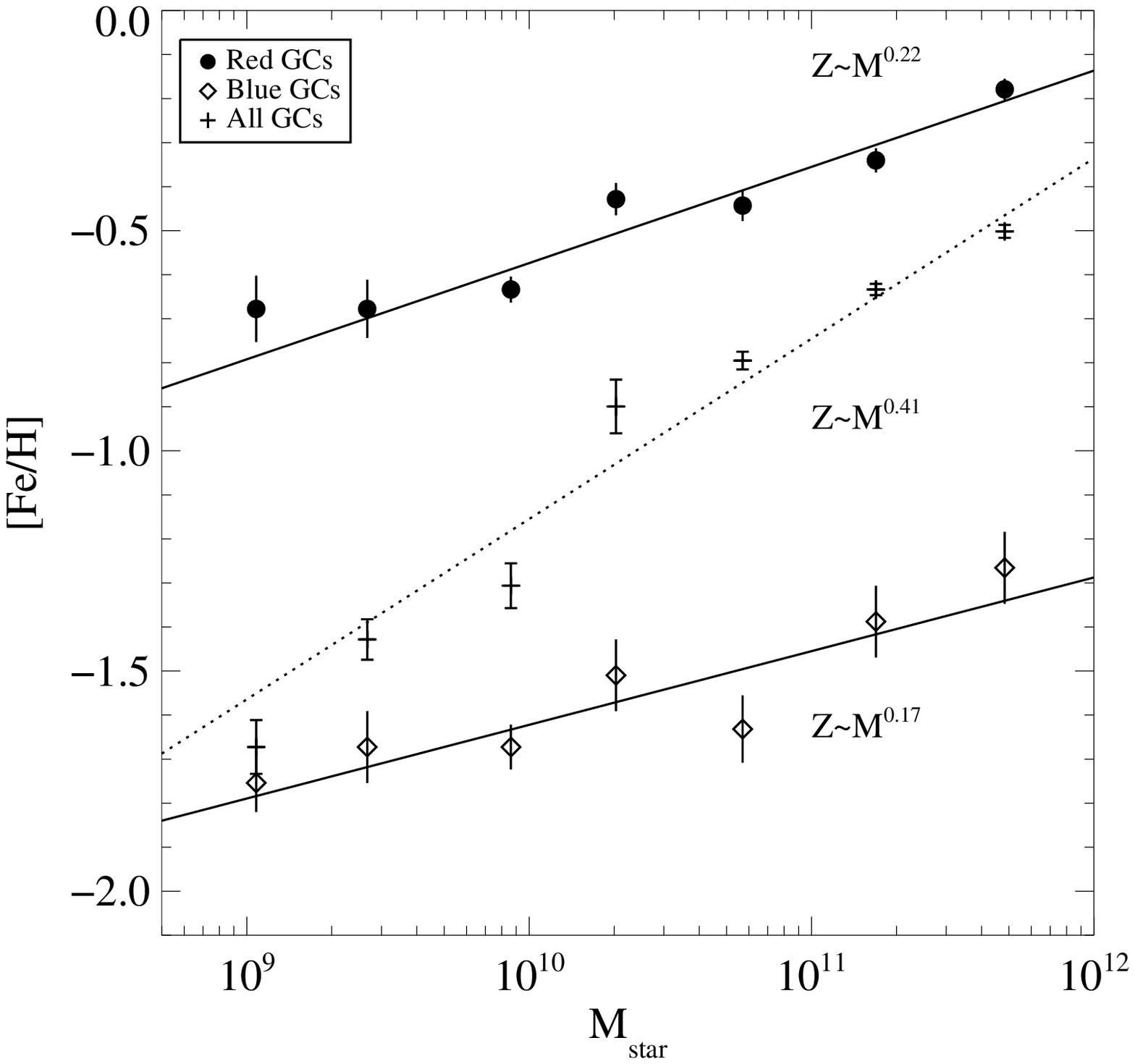}
\caption{Mean metallicities of blue, red, and total GC populations as
  a function of galaxy stellar mass.  Stellar mass has been translated
  from luminosity using an average, luminosity-weighted mass-to-light
  ratio.  Data for this plot is in Table~\ref{table:bindecomp_mass}. 
  \label{fig:binfehmass}}
\end{figure}

Figure~\ref{fig:binfehmass} shows the relationship between the
metallicity of the GC subpopulations 
and galaxy stellar mass.  Similar to the previous figure,
the metallicities of both GC subpopulations correlate strongly with
galaxy mass.  Across nearly four orders of magnitude in stellar mass, 
we find that $Z\propto M_{\star}^\beta$ where $\beta =
0.17\pm0.04,\ 0.22\pm0.03,\ {\rm and}\ 0.41\pm0.01$ for the metal-poor, metal-rich,
and total GC populations.  

The more obvious manifestation of the correlation in the metal-poor GCs 
is due to the nature of the \gz-[Fe/H] relation, which
steepens for bluer/metal-poor stellar populations.  One implication of
this is that it is a difficult task to tease out any differences between
metal-poor clusters using broadband colors, and that the apparent
universality of the population (at least in metallicity) may be in part due
to photometric errors, small numbers of GCs, or the use of a less sensitive 
color such as $V$-$I$.  Almost all previous studies on this topic have
also used linear color-metallicity relations in either $V$--$I$ or $C$--$T_1$.

Larsen \etal (2001) derive slopes between $V$--$I$ and $M_B$ 
of $-0.016\pm0.005$ and
$-0.020\pm0.008$ for the blue and red GCs, respectively.  If we convert
these to [Fe/H] using the relation of Barmby \etal (2000), which has a
slope of 4.22, then this gives $\alpha = 0.17\ {\rm and}\ 0.21$ for \
the blue and
red GCs.  As with the slopes we derived with KMM on individual galaxies,
the blue GCs have a slightly steeper, and the red GCs a slightly
shallower slope than what we finally derive with the binned sample.
However, given the differences in sample, filters, and the assumption of
a linear color-metallicity relation, the comparison can be considered
consistent.

\begin{figure}
\epsscale{1.2}
\plotone{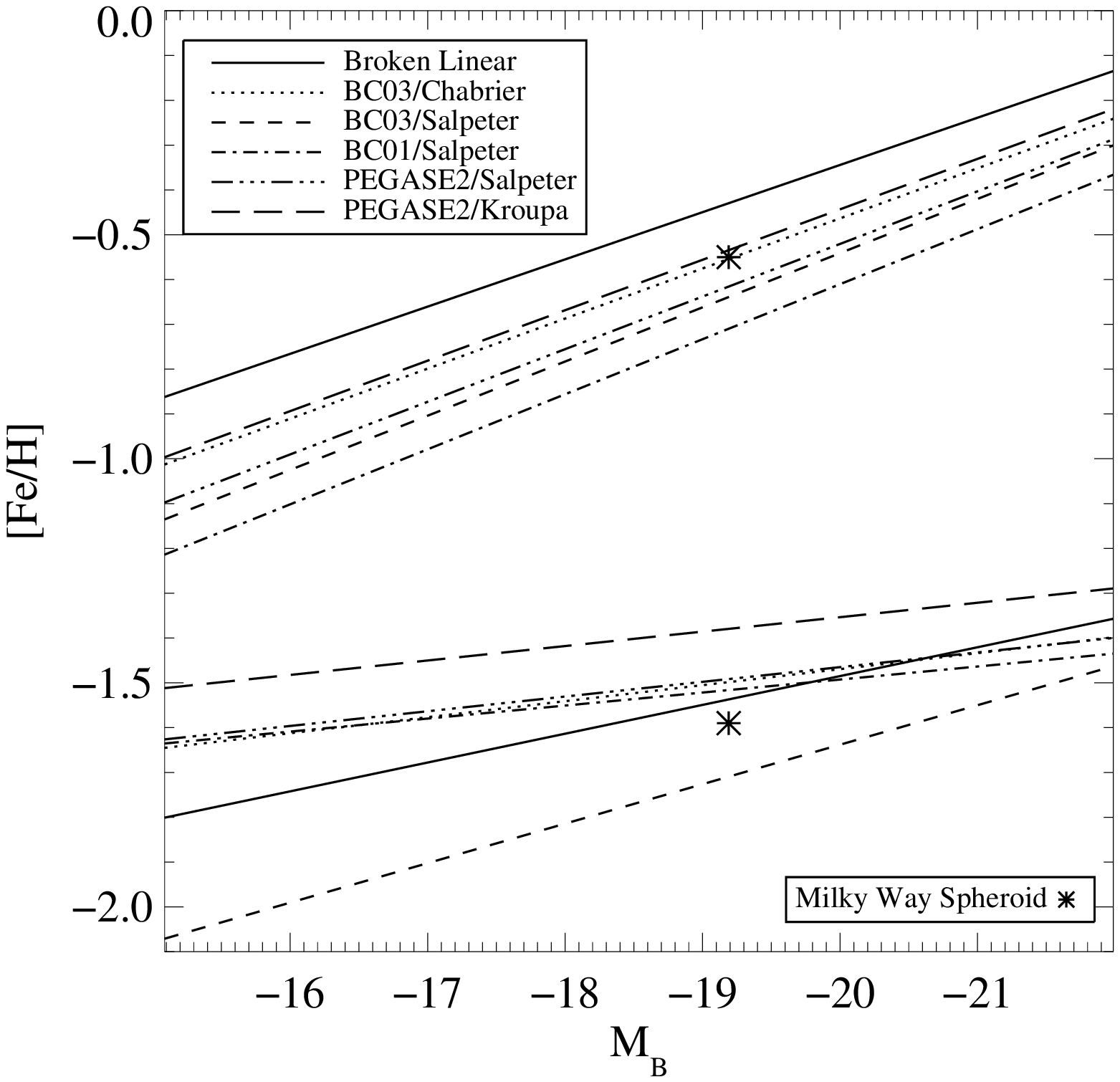}
\caption{Linear fits to the [Fe/H]-$M_B$ relations assuming
  different transformations from $(g$--$z)_0$ to [Fe/H].
  We show fits using different models with different IMFs
  as well as our empirical broken linear fit to the Galactic GC data.
  While the slope of the relation in the red GC metallicity regime is
  generally robust across different assumed transformations, the slope
  of the blue GC relation can vary by a factor of
  three. \label{fig:fituncertainty}}
\end{figure}

We emphasize, however, that the significances of
the correlations are {\it independent} of the color-metallicity relation.
It is the exact values of these proportionalities, especially the one
for the metal-poor GCs,  that are critically dependent on the adopted
\gz-[Fe/H] relation.  We illustrate this in
Figure~\ref{fig:fituncertainty}, in which we plot the various linear
fits for the $M_B$-[Fe/H] data assuming different transformations from
\gz\ to [Fe/H].  The models we use are the same as those in
Figure~\ref{fig:fehgzmodels}, and as in that Figure we linearly
interpolate the models.  While the slopes for the metal-rich
GCs are all very similar, the slopes for the metal-poor GCs can vary by
a factor of three.  However, one bit of independent reassurance is visible in
Figures~\ref{fig:binfehlum} and \ref{fig:fituncertainty}
where the large asterisk represents the
Milky Way spheroid (bulge and halo), and its associated metal-rich and
metal-poor populations of globular clusters.  We take the luminosity of
the Galactic spheroid to be $M_B=-19.19$ from \cote (1999).
That these values are close to what we might predict from
the relations derived for Virgo ellipticals is interesting in itself,
and lends credibility to our empirical color-metallicity transformation.

\subsection{The Formation and Evolution of Globular Cluster Systems}

Our ability to characterize the GC metallicity distributions across
a large range of galaxy masses provides a clearer and more precise
view of the nature of globular cluster systems.  The continuity of
GC system properties that we see paints a picture in which the
formation and evolution of GC systems shares a common mechanism across
all galaxies.  Other properties of GC systems are also either constant or
slowly varying, such as the luminosity
function, the size distribution (\jordan \etal 2005a), and the GC 
formation efficiency---globular clusters make up 0.25\% of the
baryonic mass of a galaxy (McLaughlin 1999).  This continuity mirrors 
the structural properties of the galaxies themselves, which show them to be
part of a continuously varying single family (Ferrarese \etal 2005,
Paper VI; \cote \etal 2005, Paper VIII).  

The correlations of the mean metallicities of 
metal-poor and metal-rich GCs with galaxy mass, luminosity, and color
show that the formation of both subpopulations are closely linked to
their parent galaxies.  
The nearly universal presence of a metal-poor population points to an
era of star formation that was ubiquitous in the early universe
at or shortly after reionization.  Even at that early epoch---one
associated with an epoch of ``halo formation''---the
metallicities of the forming star clusters are directly affected by
the depth of the potential well that will eventually host the GC system.
The metal-rich GCs appear to be
associated with the formation of bulges and
ellipticals---the metal-rich spheroid that defines the Hubble sequence
in the local universe.  Their numbers correlate strongly with the
mass of the spheroid, indicating that they may have formed in the same
events.  Comparisons between GC and planetary nebula
kinematics also support this view (Peng \etal 2004).  

The relationship between the mean metallicity of a stellar system and
its mass is a record of the enrichment and merging that occurred.
There is a well established mass-metallicity relationship for the
field stars of dwarf galaxies (e.g.\ Dekel \& Woo 2003) and for the gas
phase abundance of normal galaxies
(Tremonti \etal 2004).  Supernova wind feedback and its associated
loss of metals has been used to explain the $Z\propto M^{0.4}$
relationship for Local Group dwarfs (Dekel \& Silk 1986) and can
reproduce the flattening of this relationship at higher mass.
In their high-resolution simulations, Kravstov \& Gnedin (2005) find
their simulated galaxies to have $Z\propto M^{0.5}$.  
Although we know little about the epoch of halo formation,
it is possible that galaxy outflows also play a role in enriching and
triggering the formation of globular clusters (Scannapieco, Weisheit
\& Harlow 2004).
Do these relationships apply to GC systems, and should they?

Interestingly, we find $Z\propto M^{0.41}$ for the {\it total} GC systems
of our sample galaxies.  This may just be a coincidence since the
slope is driven by the fraction of metal-rich GCs and hence must
flatten at lower masses to follow the metal-poor relation.
Moreover, it is in the dwarf regime where $\alpha=0.4$ 
is seen for the field stars, not for more massive galaxies.  
However, if we assume that the
metal-poor and metal-rich GCs were formed in different events then we
might expect that their individual $\alpha$ values would reflect the
global metallicities of their host galaxies.  In the mass range that
we are concerned with ($10^9$ to $10^{12} M_\sun$), the slope of the 
$M_\star$--$Z$
relationship for galaxies is significantly shallower than
$\alpha=0.4$.  Fitting the effective yield data for the SDSS galaxies
presented in Table~4 and Figure~8 of Tremonti \etal (2004), we find
that in this mass range $Z\propto M^{0.19}$, which is quite similar the
relation for both GC subpopulations, and is also significantly
shallower than the relation seen for Local Group dwarf galaxies.
This suggests that star formation and feedback in both the epochs of halo
and metal-rich spheroid formation may not have been too different from
each other or from that observed in the present day.
In addition, we find that the Milky Way's total spheroid falls neatly on our
relations derived for Virgo ellipticals.  While there are some
dependencies between the two samples because our color-metallicity
relation was partly calibrated using Galactic GCs, the luminosity of
the spheroid is independent.  This agreement suggests that spheroids and
GC systems in
disk galaxies may form in much the same was as cluster ellipticals,
although a large census of GC systems in disk galaxies
will be necessary to explore such issues.

The slopes of the mass-metallicity
relationships are surprisingly similar for metal-rich and metal-poor GCs.  
One consequence of this is a nearly constant offset
between the metallicities of the two populations across nearly three
orders of magnitude in mass, 
$\Delta{\rm [Fe/H]}\propto M_\star^{0.05}\sim 1\ {\rm dex}$.  
If the two populations are
formed in different star forming events, this offset could point to a
characteristic enrichment that is attained between starbursts.

Recent studies of galaxies in general point to the
existence of a bimodality in galaxy properties about a
``characteristic mass'' of $3\times10^{10} M_\sun$ 
(Kauffmann \etal 2003).  Galaxies with higher
masses tend to be spheroids with old stellar populations, and those
with lower masses are likely to be star-forming disks.  This
bimodal rather than continuous distribution of galaxy properties may
have its roots in the details of gas inflow and outflow, or feedback
from active galactic nuclei (Dekel \& Birnboim 2005).  
For the GC systems, the only property that appears to change at this
mass scale is the fraction of red GCs, which is nearly constant above
this mass, and drops quickly below it.  This reinforces the idea that
the metal-rich GCs are linked to the formation of the metal-rich spheroid.
This characteristic mass is also
the regime at which the mass-metallicity relation of star forming
galaxies appears to flatten (Tremonti \etal 2004).  It is
interesting to note then, 
although perhaps not entirely surprising since our sample
consists only of early-types, that this characteristic mass does not
appear to be important for the mass-metallicity relationships of 
GC systems.  All galaxies appear to possess 
some old, GC-like stellar population, regardless of
mass, gas content, or Hubble type (see Chandar, Whitmore, \& Lee 2004, 
Olsen \etal 2004 for GCs in spiral galaxies and Seth \etal 2004 for dwarf
irregulars).  More observations are necessary to determine whether
the GC systems of disk galaxies other than the Milky Way are consistent
with the same trends we see in this study, but if they are then it would
point to a scenario where the processes that are affected by this
characteristic mass scale occur after (or independently from) the formation
of the GC system.  This is further supported by the constant offset in
metallicity between GCs and the field stars of $\sim0.8$~dex
that has been observed
across a large range of galaxy luminosity (\jordan \etal 2004c).
We speculate that this could be because 
massive star clusters form early in a star formation episode
and are less affected by the subsequent feedback-related effects that
shape the main body of the galaxy.

With larger numbers of GCs in our sample allowing us to study galaxies
in bulk, we choose to defer debates over whether a single GC
system is unimodal, bimodal, or multimodal.  Occasional claims for
multimodality have been made for specific galaxies (in particular
because of a population of intermediate-age metal-rich clusters), 
although it is very difficult to tease out such effects with single
color data alone.  In our data, we assess that VCC 881 and 798 show color
distributions that are potentially multimodal, but the occurrence is
otherwise rare or too difficult to discern.  While there are always
individual galaxies with unique GC systems, the general trends are
unmistakable.  GC color distributions are on average bimodal or asymmetric 
down to the magnitude limit of our sample, $M_B\sim-15$.  Almost all
galaxies have a population of metal-poor GCs, the mean metallicity of
which increases with galaxy mass.  These galaxies also possess some
number of more metal-rich GCs whose number fraction and mean
metallicity are also strong functions of galaxy mass.

\begin{figure}
\epsscale{1.2}
\plotone{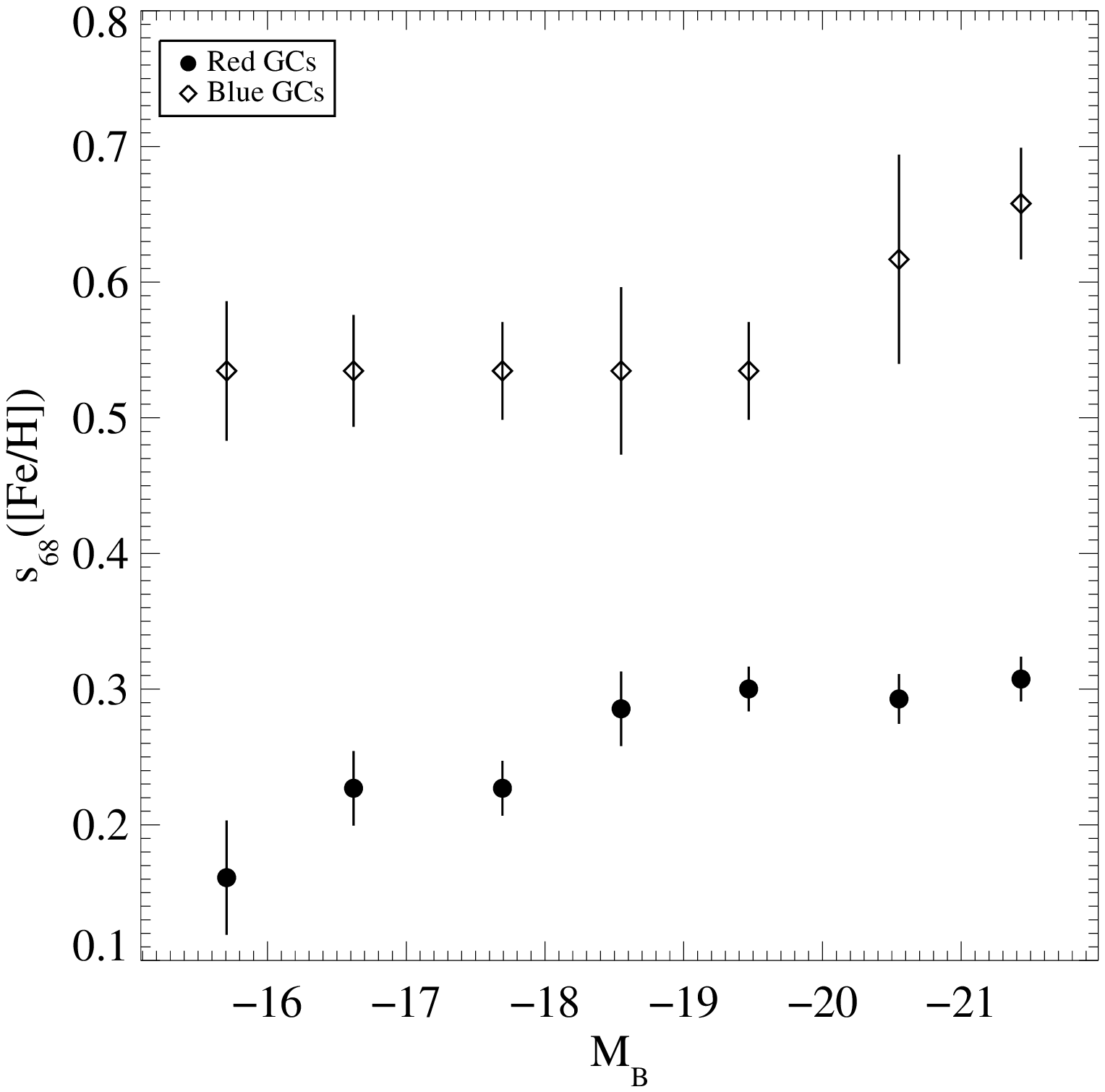}
\caption{Metallicity dispersions (68\% half-width) for the red and blue
  GC populations.  Dispersions were calculated using the color
  dispersions and transforming them to metallicity using the proper
  branch of the \gz-[Fe/H] relation.  Note that although the red GCs had
  a larger dispersion in color (Figure~\ref{fig:binsig}), the steeper
  slope of the color-metallicity relation in the blue results in the
  blue GCs having the larger dispersion in metallicity.
  \label{fig:binfehdisp}}
\end{figure}

The metallicity dispersions of GC systems increases
rapidly for more massive galaxies, and this is largely driven by the
increasing fraction of the red GCs.  When comparing the metallicity
dispersions of the red and blue GCs, the
red GCs in massive galaxies do seem to have larger dispersions
in {\it color} than the blue GCs, a factor of 1.3 in the mean.  However,
the shallower slope (i.e.\ increased metallicity
sensitivity) of the \gz--[Fe/H] relation at ${\rm [Fe/H]} >-0.8$ means that
they do not translate into larger dispersions in metallicity.
In fact, the slope of the color-metallicity relation is $2.8\pm0.5$
times steeper for metal-poor clusters, which means that it is the
{\it metal-poor GCs that have the larger dispersion in metallicity}.
This is illustrated in Figure~\ref{fig:binfehdisp}, where the median
dispersion is $0.53\pm0.07$~dex for the metal-poor GCs, 
and $\sim0.29\pm0.04$~dex
for the metal-rich GCs, where the errors reflect the formal errors in
the slope of the \gz-[Fe/H] relation.  

The dispersion in metallicity potentially tells us something about the
degree of gas mixing that occurred during galaxy formation.  A large
metallicity dispersion could be caused by prolonged
formation in a single galaxy, or also by a combination of 
star formation events that are chemically isolated from each other 
(whether spatially or temporally) and that only merge 
together after star formation has ceased.  This latter scenario could apply
to both metal-poor globular clusters that formed from isolated gaseous
fragments in the early universe, or to metal-rich clusters that form
during the hierarchical build up of the metal-rich spheroid.
It will be interesting to see whether the lack of a larger dispersion
in the metal-rich GCs, if confirmed, poses a problem for secondary
formation scenarios that require the red GCs to be the product of
multiple star forming events at moderate to low redshift, or if the
metallicity resolution is still inadequate to resolve these effects.
A narrower dispersion in metallicity also suggests that the metal-rich
population may be formed in few large starforming events where the gas is
well mixed.

Naturally, this inference is
highly dependent on the color-metallicity relation, but we point out
that none of the models shown in Figure~\ref{fig:fehgzmodels} will
produce a larger dispersion for the metal-rich GCs.  At best, the two
populations have the same dispersion.  
The Milky Way GC system does not
show such a large dispersion difference between the metal-poor and
metal-rich GCs, however, and so we will likely need the final \gz-[Fe/H]
calibration before we can verify this intriguing result.

\section{Conclusions}

Ever since the launch of $HST$, the color distributions of
extragalactic globular cluster systems have provided a unique window
onto the evolution of elliptical galaxies.  In this paper we present
the globular cluster 
color distributions in \gz\ of 100 early-type galaxies from the ACS
Virgo Cluster Survey.  As the largest sample of its kind, studied
with the increased sensitivity and spatial sampling of the $ACS/WFC$,
and with the $g_{475}$ and $z_{850}$ bandpasses, this is one of the
definitive data sets for extragalactic GC studies.  We use
carefully constructed control fields to account for galaxy specific
and spatially varying foreground and background contamination.  We
also use a new size-magnitude algorithm to produce a clean sample of
GCs.  Using both two-Gaussian models and a nonparametric
decomposition of the color distributions, we measure the
characteristics of the traditional metal-poor and metal-rich
subcomponents of the GC systems.  We also present a preliminary
transformation from \gz\ to [Fe/H] using the colors of Galactic GCs,
as well as GCs in M49 and M87.  Our main conclusions are:

\begin{enumerate}
\item While the color distributions of individual galaxies can show
  significant variations from one another, their general properties
  are consistent with {\it continuous} trends across galaxy
  luminosity, color, and mass.  Both dwarf and giant ellipticals
  appear to be subject to the same family of processes that drive the
  formation and evolution of GC systems.

\item Galaxies at all luminosities in our study, on average, appear to
  have bimodal or asymmetric GC color distributions.  {\it All}
  galaxies have a system of blue GCs with an average color around
  $g$--$z \sim 0.9$ and a red component whose fraction varies with
  the luminosity or color of the galaxy.  When decomposed
  into blue and red GC subpopulations, the red GCs on average
  compose $\sim15\%$
  of the GCs in the faintest and bluest galaxies in our sample 
  ($M_B=-16$, $g$--$z=1.0$), 
  and up to 60\% of the GCs the brightest
  and reddest galaxies ($M_B=-21$, $g$--$z=1.5$).

\item The colors of the blue and red GC subpopulations correlate with
  the luminosity and color of the host galaxy.  In \gz, the slope
  determined for the red GCs is 4.6 times steeper than for the blue
  GCs.

\item The color and metallicity widths of the entire GC system
  increases with 
  galaxy luminosity and color.  The width of the red
  population is larger than that of the blue population {\it in color},
  but because of the decreased sensitivity
  of color to metallicity in the blue, the widths of the red peak 
  {\it in metallicity} are consistent with being {\it smaller} than
  those of the blue peak.  

\item Using new Galactic GC imaging combined with M49 and M87 GCs that
  have spectroscopic metallicities, we have defined a new preliminary
  color-metallicity relation for globular clusters.  We find that this
  relation shows clear nonlinearity and parametrize the relation with
  a broken line with a break point at \gz=1.05, [Fe/H]$=-0.81$.  The
  steepening of the relation at low metallicity is critical to
  the proper interpretation of metal-poor GC colors.
  Current evolutionary synthesis models produce colors for 13~Gyr
  SSPs that are roughly consistent with our color-metallicity
  relation, but show large variation for $g$--$z<1.0$.

\item The metallicities of the metal-rich and metal-poor GC
  subpopulations both correlate with the luminosity of the host
  galaxy, with the slope for the metal-rich GCs 1.7 times steeper.
  The derived slopes are not nearly as disparate in metallicity as
  they are in color because of the steeper color-metallicity relation
  for metal-poor GCs.  However, uncertainty in the \gz--[Fe/H] relation
  in the blue can cause this slope to change by a factor of three.
  We find that 
  $Z\propto L^{0.16\pm0.04}$ for metal-poor GCs, 
  $Z\propto L^{0.25\pm0.03}$ for metal-rich GCs , and
  $Z\propto L^{0.52\pm0.02}$ for the total GC population.

\item We use the optical-infrared colors of the ACSVCS galaxies to
  derive crude stellar masses and find that the metallicities of the
  GC populations also correlate with the stellar mass of the host
  galaxy.  The ratio of the slopes is smaller again, with the
  slope for the metal-rich GCs 1.4 times steeper.  We find 
  $Z\propto M_\star^{0.17\pm0.04}$ for metal-poor GCs,  
  $Z\propto M_\star^{0.22\pm0.03}$ for metal-rich GCs, and 
  $Z\propto M_\star^{0.41\pm0.01}$ for the total GC population.

\item The small difference in slope in the $M_\star$--$Z$ relation means that
  the metallicity difference between the metal-poor and metal-rich GC
  subpopulations is a nearly constant $\Delta{\rm [Fe/H]}\sim 1\ {\rm
  dex}$, varying slowly with stellar mass as $M_\star^{0.05}$.

\end{enumerate}

As part of the ACSVCS, we have also obtained long-slit spectroscopy of
the entire galaxy sample.  In a subsequent paper, we will present the
properties of the GC color distributions as a function of galaxy
metallicity, alpha-enhancement, and total dynamical mass.  A future
paper will also address these results in the context of detailed
modeling of host galaxy merging histories.

\acknowledgments

EP thanks C.\ Tremonti for useful discussions on the mass-metallicity
relation.  Support for program GO-9401 was provided through a grant from the
Space Telescope Science Institute, which is operated by the
Association for Research in Astronomy, Inc., under NASA contract NAS5-26555.
Partial support for this work was provided by NASA LTSA grant
NAG5-11714 to PC.  M. J. W. acknowledges support
through NSF grant AST 02-05960.
This research has made use of the NASA/IPAC Extragalactic Database
(NED) which is operated by the Jet Propulsion Laboratory, California
Institute of Technology, under contract with the National Aeronautics
and Space Administration. 
This publication makes use of data products from the Two Micron All Sky
Survey, which is a joint project of the University of Massachusetts and
the Infrared Processing and Analysis Center/California Institute of
Technology, funded by the National Aeronautics and Space Administration
and the National Science Foundation. 



Facilities: HST(ACS), CTIO.

\clearpage

\begin{deluxetable}{lcrr}
\tablewidth{0pt}
\tablecaption{Linear fits to Red and Blue GC components of individual galaxies versus $M_B$\label{table:clfittable}}
\tablehead{
\colhead{} & 
\colhead{} & 
\colhead{$\langle g-z \rangle = a + b\times M_B$} & 
\colhead{} \\
\colhead{} & 
\colhead{p-value} & \colhead{$a$} & \colhead{$b$} 
}
\startdata
Red GCs & $p<0.05$ & $ 0.629\pm 0.082$ & $-0.036\pm 0.004$ \\
 & $0<p<1$ & $ 0.503\pm 0.061$ & $-0.042\pm 0.003$ \\
Blue GCs & $p<0.05$ & $ 0.424\pm 0.063$ & $-0.026\pm 0.003$ \\
 & $0<p<1$ & $ 0.462\pm 0.046$ & $-0.024\pm 0.002$ \\
\enddata
\end{deluxetable}

\begin{deluxetable}{lccccccccccccccc}
\tablewidth{0pt}
\tabletypesize{\tiny}
\tablecaption{Two-Component Decompositions for GC Color Distributions, Binned by $M_B$ \label{table:bindecomp_mag}}
\tablehead{
\colhead{$M_B$} & 
\colhead{$\mu_b$} & 
\colhead{$\mu_r$} & 
\colhead{$s_{68,b}$} & 
\colhead{$s_{68,r}$} & 
\colhead{$f_{red}$} & 
\colhead{$\langle (g$--$z)_{host}\rangle$} & 
\colhead{$N_{gc}$} & 
\colhead{$\mu_1$} & 
\colhead{$\sigma_1$} \\ 
\colhead{(1)} & 
\colhead{(2)} & 
\colhead{(3)} & 
\colhead{(4)} & 
\colhead{(5)} & 
\colhead{(6)} & 
\colhead{(7)} & 
\colhead{(8)} & 
\colhead{(9)} & 
\colhead{(10)} 
}
\startdata
-21.4 & $0.95\pm0.01$ & $1.40\pm0.01$ & $0.13\pm0.01$ & $0.17\pm0.01$ & $0.63\pm0.03$ & 1.50 & 3214 & $1.26\pm0.01$ & $0.30\pm0.01$ \\
-20.5 & $0.93\pm0.02$ & $1.32\pm0.02$ & $0.12\pm0.01$ & $0.16\pm0.01$ & $0.56\pm0.06$ & 1.47 & 3122 & $1.15\pm0.01$ & $0.26\pm0.01$ \\
-19.5 & $0.91\pm0.01$ & $1.28\pm0.02$ & $0.10\pm0.01$ & $0.16\pm0.01$ & $0.52\pm0.04$ & 1.44 & 1692 & $1.08\pm0.01$ & $0.26\pm0.01$ \\
-18.5 & $0.89\pm0.02$ & $1.24\pm0.03$ & $0.10\pm0.01$ & $0.16\pm0.01$ & $0.51\pm0.06$ & 1.38 & 1088 & $1.04\pm0.01$ & $0.24\pm0.01$ \\
-17.7 & $0.89\pm0.01$ & $1.16\pm0.02$ & $0.10\pm0.01$ & $0.12\pm0.01$ & $0.33\pm0.04$ & 1.36 &  644 & $0.96\pm0.01$ & $0.18\pm0.01$ \\
-16.6 & $0.87\pm0.01$ & $1.13\pm0.03$ & $0.10\pm0.01$ & $0.12\pm0.01$ & $0.29\pm0.06$ & 1.14 &  505 & $0.92\pm0.01$ & $0.17\pm0.01$ \\
-15.7 & $0.88\pm0.01$ & $1.12\pm0.04$ & $0.10\pm0.01$ & $0.09\pm0.02$ & $0.15\pm0.06$ & 1.14 &  233 & $0.89\pm0.01$ & $0.14\pm0.01$ \\
\enddata
\tablenotetext{1}{Mean galaxy magnitude in bin}
\tablenotetext{2,3}{Median colors of blue and red GC components}
\tablenotetext{4,5}{68\% half-widths of blue and red GC components}
\tablenotetext{6}{Fraction of clusters in red component}
\tablenotetext{7}{Mean color of galaxies in this bin}
\tablenotetext{8}{Number of GCs}
\tablenotetext{9}{Median of entire GC distribution}
\tablenotetext{10}{68\% half-width of entire GC distribution}
\end{deluxetable}

\begin{deluxetable}{lccccccccccccccc}
\tablewidth{0pt}
\tabletypesize{\tiny}
\tablecaption{Two-Component Decompositions for GC Color Distributions, Binned by galaxy $(g-z)$ \label{table:bindecomp_col}}
\tablehead{
\colhead{$(g$--$z)_0$} & 
\colhead{$\mu_b$} & 
\colhead{$\mu_r$} & 
\colhead{$s_{68,b}$} & 
\colhead{$s_{68,r}$} & 
\colhead{$f_{red}$} & 
\colhead{$N_{gc}$} & 
\colhead{$\mu_1$} & 
\colhead{$\sigma_1$} \\ 
\colhead{(1)} & 
\colhead{(2)} & 
\colhead{(3)} & 
\colhead{(4)} & 
\colhead{(5)} & 
\colhead{(6)} & 
\colhead{(7)} & 
\colhead{(8)} & 
\colhead{(9)} 
}
\startdata
1.52 & $0.95\pm0.02$ & $1.40\pm0.02$ & $0.13\pm0.01$ & $0.16\pm0.01$ & $0.55\pm0.05$ & 2064 & $1.20\pm0.01$ & $0.30\pm0.01$ \\
1.47 & $0.91\pm0.00$ & $1.32\pm0.01$ & $0.10\pm0.01$ & $0.18\pm0.01$ & $0.58\pm0.02$ & 6375 & $1.13\pm0.01$ & $0.28\pm0.01$ \\
1.37 & $0.91\pm0.02$ & $1.23\pm0.02$ & $0.12\pm0.01$ & $0.15\pm0.01$ & $0.47\pm0.05$ & 1439 & $1.04\pm0.01$ & $0.23\pm0.01$ \\
1.26 & $0.88\pm0.02$ & $1.16\pm0.05$ & $0.10\pm0.01$ & $0.13\pm0.02$ & $0.31\pm0.08$ &  518 & $0.94\pm0.01$ & $0.18\pm0.01$ \\
1.14 & $0.89\pm0.01$ & $1.12\pm0.03$ & $0.11\pm0.01$ & $0.09\pm0.01$ & $0.18\pm0.05$ &  393 & $0.92\pm0.01$ & $0.14\pm0.01$ \\
1.02 & $0.87\pm0.01$ & $1.14\pm0.06$ & $0.13\pm0.02$ & $0.10\pm0.04$ & $0.14\pm0.07$ &   99 & $0.88\pm0.02$ & $0.15\pm0.02$ \\
\enddata
\tablenotetext{1}{Mean galaxy color in bin}
\tablenotetext{2,3}{Median colors of blue and red GC components}
\tablenotetext{4,5}{68\% half-widths of blue and red GC components}
\tablenotetext{6}{Fraction of clusters in red component}
\tablenotetext{7}{Number of GCs}
\tablenotetext{8}{Median of entire GC distribution}
\tablenotetext{9}{68\% half-width of entire GC distribution}
\end{deluxetable}

\begin{deluxetable}{cclrr}
\tablewidth{0pt}
\tablecaption{Linear fits to Red and Blue GC components of versus Galaxy Properties\label{table:binfit}}
\tablehead{
\colhead{$X$} & 
\colhead{$Y$} & 
\colhead{} & 
\colhead{$Y = a + bX$} & 
\colhead{} \\
\colhead{Galaxy Property} & 
\colhead{GCS Property} & 
\colhead{GCs} & \colhead{$a$} & \colhead{$b$} 
}
\startdata
$M_B$ & $(g$--$z)$ & Red & $ 0.163\pm 0.006$ & $-0.058\pm 0.001$ \\
 & & Blue & $ 0.668\pm 0.046$ & $-0.013\pm 0.003$ \\
$(g$--$z)$ & $(g$--$z)$ & Red & $ 0.405\pm 0.091$ & $ 0.622\pm 0.063$ \\
 & & Blue & $ 0.795\pm 0.033$ & $ 0.078\pm 0.024$ \\
$M_B$ & [Fe/H] & Red & $-2.452\pm 0.021$ & $-0.105\pm 0.013$ \\
 & & Blue & $-2.771\pm 0.232$ & $-0.064\pm 0.015$ \\
 & & All & $-4.854\pm 0.124$ & $-0.206\pm 0.006$ \\
$log_{10}(M_{\star})$ & [Fe/H] & Red & $-2.759\pm 0.032$ & $ 0.219\pm 0.028$ \\
 & & Blue & $-3.296\pm 0.328$ & $ 0.167\pm 0.039$ \\
 & & All & $-5.250\pm 0.156$ & $ 0.409\pm 0.014$ \\
\enddata
\end{deluxetable}

\begin{deluxetable}{lccccccccccccccc}
\tablewidth{0pt}
\tabletypesize{\tiny}
\tablecaption{Two-Component Decompositions for GC Color Distributions, Binned by Stellar Mass \label{table:bindecomp_mass}}
\tablehead{
\colhead{$M_\star (M_\sun)$} & 
\colhead{$\mu_b$} & 
\colhead{$\mu_r$} & 
\colhead{$s_{68,b}$} & 
\colhead{$s_{68,r}$} & 
\colhead{$f_{red}$} & 
\colhead{$\langle (g$--$z)_{host}\rangle$} & 
\colhead{$N_{gc}$} & 
\colhead{$\mu_1$} & 
\colhead{$\sigma_1$} \\ 
\colhead{(1)} & 
\colhead{(2)} & 
\colhead{(3)} & 
\colhead{(4)} & 
\colhead{(5)} & 
\colhead{(6)} & 
\colhead{(7)} & 
\colhead{(8)} & 
\colhead{(9)} & 
\colhead{(10)} 
}
\startdata
  4.8E+11 & $0.96\pm0.02$ & $1.40\pm0.01$ & $0.14\pm0.01$ & $0.16\pm0.01$ & $0.58\pm0.04$ & 1.50 & 3558 & $1.23\pm0.01$ & $0.30\pm0.01$ \\
  1.7E+11 & $0.94\pm0.02$ & $1.32\pm0.01$ & $0.12\pm0.01$ & $0.16\pm0.01$ & $0.57\pm0.04$ & 1.48 & 2789 & $1.16\pm0.01$ & $0.26\pm0.01$ \\
  5.7E+10 & $0.89\pm0.01$ & $1.26\pm0.02$ & $0.10\pm0.01$ & $0.16\pm0.01$ & $0.54\pm0.04$ & 1.46 & 1326 & $1.07\pm0.01$ & $0.25\pm0.01$ \\
  2.0E+10 & $0.92\pm0.02$ & $1.27\pm0.02$ & $0.12\pm0.01$ & $0.15\pm0.01$ & $0.44\pm0.05$ & 1.39 & 1061 & $1.04\pm0.01$ & $0.24\pm0.01$ \\
  8.6E+09 & $0.88\pm0.01$ & $1.16\pm0.02$ & $0.10\pm0.01$ & $0.13\pm0.01$ & $0.37\pm0.04$ & 1.33 &  528 & $0.96\pm0.01$ & $0.18\pm0.01$ \\
  2.7E+09 & $0.88\pm0.02$ & $1.13\pm0.04$ & $0.10\pm0.01$ & $0.12\pm0.01$ & $0.32\pm0.07$ & 1.20 &  487 & $0.93\pm0.01$ & $0.16\pm0.01$ \\
  1.1E+09 & $0.87\pm0.01$ & $1.13\pm0.04$ & $0.11\pm0.01$ & $0.09\pm0.03$ & $0.16\pm0.06$ & 1.11 &  268 & $0.88\pm0.01$ & $0.15\pm0.01$ \\
\enddata
\tablenotetext{1}{Mean galaxy stellar mass in bin}
\tablenotetext{2,3}{Median colors of blue and red GC components}
\tablenotetext{4,5}{68\% half-widths of blue and red GC components}
\tablenotetext{6}{Fraction of clusters in red component}
\tablenotetext{7}{Mean color of galaxies in this bin}
\tablenotetext{8}{Number of GCs}
\tablenotetext{9}{Median of entire GC distribution}
\tablenotetext{10}{68\% half-width of entire GC distribution}
\end{deluxetable}







\end{document}